\documentclass[preprint,12pt,authoryear]{elsarticle}
\usepackage{amsmath}
\usepackage{amssymb}

\usepackage{lineno}
\usepackage{booktabs}
\usepackage{array}
\usepackage{tabulary}
\usepackage[section]{placeins}
\usepackage{titlesec}
\usepackage{longtable}
\usepackage{cleveref}

\bibpunct[:~]{(}{)}{;}{a}{}{;}

\journal{Neural Networks}
\begin{document}

\begin{frontmatter}
\title{Connectivity Inference from Neural Recording Data: Challenges, Mathematical Bases and Research Directions}
\author[oist,araya]{Ildefons Magrans de Abril}
\author[naist]{Junichiro Yoshimoto}
\author[oist]{Kenji Doya}
\address[oist]{Okinawa Institute of Science and Technology Graduate University}
\address[naist]{Nara Advanced Institute of Science and Technology}
\address[araya]{ARAYA, Inc., Tokyo, Japan}
\begin{abstract}
This article presents a review of computational methods for connectivity inference from neural activity data derived from multi-electrode recordings or fluorescence imaging. We first identify biophysical and technical challenges in connectivity inference along the data processing pipeline. We then review connectivity inference methods based on two major mathematical foundations, namely, descriptive model-free approaches and generative model-based approaches. We investigate representative studies in both categories and clarify which challenges have been addressed by which method. We further identify critical open issues and possible research directions. 
\end{abstract}
\end{frontmatter}

\section{Introduction}
\label{introduction}

Understanding the operational principles of neural circuits is a major goal of recent international brain science programs, such as the BRAIN Initiative in the U.S. \citep{insel2013nih,Martin2016570}, the Human Brain Project in the E.U. \citep{markram2012human, Amunts2016574}, and the Brain/MINDS program in Japan \citep{Okano2015nsr, Okano2016582}. A common emphasis in these programs is utilization of high-throughput, systematic data acquisition and advanced computational technologies. The aim of this paper is to present a systematic review of computational methods for inferring neural connectivity from high-dimensional neural activity recording data, such as multiple electrode arrays and calcium fluorescence imaging. 

Why do we need to infer neural connectivity? High-dimensional neural recording data tell us a lot about information representation in the brain through correlation or decoding analyses with relevant sensory, motor, or cognitive signals. However, in order to understand the operational principles of the brain, it is important to identify the circuit mechanisms that encode and transform information, such as extraction of sensory features and production of motor action patterns \citep{Churchland1992cb}. Knowing the wiring diagram of neuronal circuits is critical in explaining how such representations can be produced, predicting how the network would behave in a novel situation, and extracting the brain's algorithms for technical applications \citep{sporns2005human}.

The network of the brain can be analyzed at various spatial scales \citep{Gerstner2014nd}. At the macroscopic level, there are more than a hundred of anatomical brain areas and connection structure across those areas give us an understanding of the overall processing architecture of the brain. At the mesoscopic level, the connections of neurons within each local area, as well as their projections to other areas, are characterized in order to understand the computational mechanisms of neural circuits. At the microscopic level, the location and features of synapses on the dendritic arbors of each neuron are analyzed to understand the operational mechanisms of single neurons.

This review focuses on the mesoscopic level, inferring the connections between neurons in local circuits from neural activity recording data by multi-electrode recordings or fluorescence imaging. Connectivity inference from anatomical data, such as diffusion MRI at the macroscopic level, tracer injection at the mesoscopic level, and serial electron microscopy data at the microscopic data, are beyond the scope of this review. Some of the methods, especially those of model-free approaches, may also be applicable to connectivity inference from functional MRI data at the macroscopic level.

%This review focuses on the mesoscopic level, inferring the connections between neurons in local circuits, utilizing multi-electrode recordings or fluorescence imaging data. Connectivity inference at the macroscopic level, investigated using MRI data, and the microscopic level using electron microscopy data, are beyond the scope of this review, but some of the methods, especially those of model-free approaches, may also be applicable to connectivity inference at macroscopic level.

This paper presents an overview of a variety of challenges in network inference and different mathematical approaches to address those challenges. We first review the data processing pipeline of connectivity analysis and identify both biophysical and computational difficulties. From mathematical view point, we classify connectivity inference methods broadly into descriptive, model-free approaches and generative, model-based approaches and explain representative methods in each category. We then examine which methods offer solutions for specific challenges and identify open issues and important research directions.

There have been several recent reviews on specific mathematical frameworks in network connectivity inference, such as the Bayesian approaches \citep{chen2013overview} and the maximum entropy method \citep{yeh2010maximum,roudi2015multi}. Some reviews focus on macroscopic connectivity analysis using MRI
\citep{friston2011functional,lang2012brain,sporns2012discovering,sporns2013human}.
The reports from the First Neural Connectomics Challenge (http://connectomics.chalearn.org)
reviews the top-ranked methods for connectivity inference from calcium imaging data \citep{orlandifirst,guyon2014design}.

It is important to distinguish several types of connectivity that have been addressed previously \citep{ aertsen1989dynamics, friston2011functional, valdes2011effective}.
\textit{Functional} connectivity (FC) is defined as statistical dependence among measurements of neuronal activity. It can be computed using correlation or other model-free methods (see section \ref{modelfreemethods}).
\textit{Effective} connectivity (EC) characterizes the direct influence exercised between neuron pairs after discounting any indirect effects. EC is usually computed by optimizing the parameters of a model that is assumed to have generated the observed data (see section \ref{modelbasedmethods}).
\textit{Anatomical} connectivity signifies existence of actual synapses, either excitatory or inhibitory. Even if there is an anatomical connection between neurons, the connection may not be detected when, for example, the source neuron is inactive or the recipient neuron is irresponsive due to strong inhibition by other neurons. Detection of functional or effective connectivity does not warrant the existence of anatomical connectivity. An example would be a false positive connection inferred between two neurons that receive common inputs from a third neuron.

%An example would be two neurons that receive common inputs from a third neuron. 

%THIS GOES TO APPARENT CONNECTIVITY DISCUSSION: does not specify the abilities to distinguish a direct activation as well as to identify the directionality of the connection are important capabilities of our inference methods. however, the definition of functional and effective connection does not impose any of these two features

Neural connectivity can be described at difference levels of detail: existence, direction, sign, magnitude, and temporal dynamics. Which level of description is most reliable and useful depends on constraints on the instrumentation and the amount of data available. 
One aim of this paper is to clarify how those constraints affect the choice of inference and the validation methods for a given application. 

Although it is beyond the scope of this review, graph theoretical analysis of the inferred network can play a key role in understanding the interplay between the brain structure and its function \citep{bullmore2009complex}.  Such graph theoretic characterization includes clustering and connectivity degree distribution \citep{shimono2014functional,bonifazi2009gabaergic,yu2008small}. These abstract metrics facilitate comparison of the structure of diverse neural populations. For example, \cite{hu2016feedback} proposed a method to relate the network statistics of connectivity of linear point processes or Hawkes models (see section \ref{sec:hawkes}) to its function.

%[McIntosh: discusses the question of how from neurophisiology we can explain how the brain produces human mental functions]
%[Park and Friston:review paper about what we know about the relationship between function structure in the brain: many to one, hierarchical, local circuit context dependent]

\section{Data Processing Pipeline} 
\label{pipeline}

\begin{figure}[h!]
  \centering
  \includegraphics[width=0.80\textwidth]{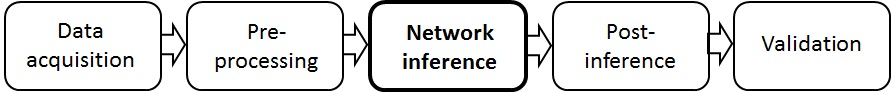}
  {\caption{Generic data processing pipeline, starting with data acquisition and continuing with data pre-processing, network inference, post-processing, and validation of results.}
  \label{fig:pipe}}
\end{figure}

This section presents a generic data processing pipeline, starting with data acquisition and continuing with data pre-processing, network inference, post-processing, and validation of results.
Figure~\ref{fig:pipe} shows the overall view of the experimental setup.

\subsection{Data Acquisition}
\label{recording}

Multiple electrodes and fluorescence imaging are two major methods for making high-dimensional measurements of single neuronal activities.

\subsubsection{Multiple electrode recording}

For \textit{in vivo} multiple neural recordings, the most commonly used device is the so-called ``tetrode,'' a bundle of four wire electrodes. By implanting tens of tetrodes and applying a spike sorting method, it is possible to record hundreds of neurons simultaneously~\citep{Jog2002,Buzsaki2004nns}.
Linear or matrix arrays of electrodes, often using semiconductor technologies, are also used for recording hundreds of neurons near the cortical surface~\citep{Fujishiro2014}.
For brain tissue slices and cultures, electrode grids patterned on a plate enable recording of hundreds of neural activities \citep{fujisawa2008behavior, shimono2014functional}. In such \textit{in vitro} experiments, it is also possible to use intracellular glass electrodes to record sub-threshold membrane potentials of selected neurons in a population~\citep{brette2012handbookchapter3}.

\subsubsection{Optical imaging}

Optical imaging allows activity data gathering from hundreds to thousands of neurons simultaneously, using voltage sensitive dyes (VSDs) or genetically encoded calcium indicators (GECIs). Chemical VSDs offer high temporal resolution, but lack the cell type-specificity that GECIs offer. However, genetically encoded voltage indicators (GEVIs) are in active development \citep{Yang2016}.
Recently GECIs have become increasingly popular because they can be expressed under the control of cell-type specific promoters \citep{Looger2011}. The weaknesses of GECIs has been their slow temporal response and low sensitivity, but recent GECIs have begun to achieve time constants on the order of ten milliseconds, making detection of single spikes feasiible \citep{Pnevmatikakis2016}. 

Fast CCD cameras and confocal or two-photon laser scanning microscopes are commonly used with GECIs. CCD cameras enable simultaneous imaging of all neurons in the focal plane, allowing recording of as many as a thousand frames per second, but measurements are limited to neurons near the surface of the tissue. 
%CCD cameras are mostly used with slice preparations. Laser scanning microscopes enable access to neurons deeper in tissues. 
Two-photon microscopes use infrared light, which is less subject to refraction and excites fluorescent molecules only at the focal point, allowing recording of neurons several hundreds of microns beneath the surface \citep{lutcke2011two,Dombeck2009}. Most recently, head-mount miniature microscopes using gradient index (GRIN) rod lenses have allowed access to deep neural structures, such as the hippocampus of awake behaving animals \citep{Ziv2013}.

\subsection{Pre-processing}
\label{preprocessing}

%Data obtained with multiple electrodes consist of multidimensional time-series of voltage recordings, while data obtained with fluorescence imaging consist of image sequences. 
Pre-processing steps depend on the recording methods and the specific input requirements of the network inference method.

In multiple electrode recording, each electrode can receive signals from multiple neurons and also the signal from the same neuron can be detected by multiple electrodes. Spike sorting algorithms are used to identify the spikes from each neuron by applying principal component analysis, independent component analysis, and any biophysical knowledge of spike shapes and intervals \citep{shimono2014functional,mahmud2015qspike}.

%Optical imaging has the capacity to simultaneously monitor thousands of neurons with sampling frequencies from ${1\,Hz}$ to ${1\,kHz}$ \citep{lutcke2011two}.
The task in optically imaged data pre-processing is to transform an image sequence into a multi-dimensional time-series of neural activities.
Pre-processing steps for optical imaging are: i) image segmentation to identify regions  corresponding to each neuron,
ii) extraction of the fluorescence trace for each neuron 
and iii) spike inference \citep{Pnevmatikakis2016}. 
These pre-processing operations have to deal with light scattering, motion artifacts, and slow fluorescence changes with respect to the underlying membrane potential.

\subsection{Connectivity inference}
\label{inference}

Connectivity inference methods can be largely classified into two classes. Descriptive model-free methods are based on descriptive statistics without assuming any particular process that generated the data. On the other hand, generative model-based methods assume a certain mathematical model that generates the data and infer the parameters and structure of the model. We will explain those methods in detail in the subsequent sections. 

Most connectivity inference methods require time-series of spikes from each neuron in the population. 
In some studies, spike inference and connectivity inference are performed using an integrated optimization algorithm directly from time series of fluorescence \citep{Mishchenko2011aas}. Some other studies use fluorescence signals directly for model-free analysis of connectivity without a explicit spike inference mechanism \citep{veeriah2015deep}. 

\subsection{Post-processing}
\label{postprocessing}

After the connectivity matrix is obtained by applying any of the connectivity inference methods, it is often useful to perform post-processing to achieve a biologically realistic result.
For example, inference methods that consider only simultaneous activities yield only symmetric connectivity matrices. 
A heuristic method to determine the direction is to use the average temporal order of activations of two neurons \citep{sutera}.
Another issue is that the inferred connectivity can be a mix of direct causal effects between neuron pairs and indirect effects through other neurons.
One way to address this problem is to use matrix deconvolution \citep{feizi2013network,barzel2013network,magrans}, as described in section \ref{modelfree_studies2}.

Furthermore, employment of a network inference method depends on the choice of parameters. It is possible to improve both robustness and accuracy of the connectivity matrix by combining several matrices computed using different parameters and/or different inference methods \citep{sutera,magrans}. 
   
\subsection{Validation}
\label{validation}

In order to validate the connectivity inference method itself, the use of synthetic data from a network model with a known connection matrix is the first step. For estimating connectivity of real biological neural networks, validation is more challenging, as anatomical or physiological verification of all pair-wise synaptic connectivity is extremely laborious.

\subsubsection{Synthetic data}
\label{syntheticdatasource}
The strength of synthetic data for validation is that we have all information about simulated connectivity and other biophysical parameters of the model. 
Therefore, we can use standard error metrics to measure the similarity between the inferred connectivity matrix and the one used for data generation.

When the final objective is to infer connection weights, one difficulty with many methods is that they deliver estimated connectivity values with different scales.
The relative mean squared error
\begin{equation}
  \label{eqn: relativermse}
  relative\,MSE =  \frac {min_{\alpha}\sum_{ij}\mid W_{ij} - \alpha \hat{W_{ij}}\mid^2}{\sum_{ij} \mid W_{ij} \mid ^2},
 \end{equation} 
 where $\hat{W_{ij}}$ is the estimate of $W_{ij}$,
 can alleviate the scaling problem \citep{fletcher2014scalable}.
  
If the objective is to infer the graphical structure of the network, what matters is the binary existence/nonexistence of connections. The Area Under the Curve of the Receiver-Operator Characteristic (AUROC) is a popular performance metric used in such a case \citep{guyon2014design,garofalo2009evaluation,stetter2012model}. 
The ROC Curve describes the relationship between the False Positive (FP) Rate ($\frac{FP}{FP+TN}$) or Fall-out and the True Positive (TP) Rate ($\frac{TP}{TP+FN}$) or Recall at different thresholds.
The perfect classifier has an AUROC of 1, while a random classifier has an AUROC of 0.5.
However, this metric may overestimate the performance in highly biased data-sets \citep{schrynemackers2013protocols}. The Area under the Precision Recall Curve (AUPRC) was proposed as an alternative measure to improve validation accuracy of sparsely connected neural networks \citep{orlandifirst,sutera}. Specifically, it is the area below the curve that describes the relationship between the Precision ($\frac{TP}{TP+FP}$) versus the Recall ($\frac{TP}{TP+FN}$) at different thresholds. 

In order to find the best trade-off between $TP$s and $FP$s, \cite{garofalo2009evaluation} defined the Positive Precision Curve (PPC),  which describes the relationship between the True False Sum ($TFS=TP+FP$) and the True False Ratio ($TFR=\frac{TP-FP}{TFS+TN}$). 
The peak of this curve can be used to extract a binary connectivity map from a weight matrix inferred by a network inference method. 
To assess the required recording duration and bin size to achieve a desired reconstruction performance, \cite{ito2011extending}
proposed the use of the curves of the TPR at a fixed FPR as a function of different recording durations and bin sizes.

\subsubsection{Real data}
\label{realdatasource}
%Our real goal is to accurately infer the connectivity structure and other biophysical parameters in real neural microcircuits. Under this scenario, 
In real neural recording, although the ground truth of the connectivy is not generally available, we can assess the quality of inferred connections using statistical significance testing and cross validation. 
Significance testing is used to accept or reject the null hypothesis ($H_0$) that a connection between two neurons does not exist \citep{lizier2014jidt}. To approximate $H_0$, we can run our network inference method on many surrogate time-series created by perturbing the training time-series such that it destroys the connectivity information \citep{fujisawa2008behavior, lizier2011multivariate, shimono2014functional, oba2016empirical}. Then, the test itself consists of computing the probability that the inferred connectivity value is generated uniquely by chance. However, sometimes an accept/reject test is not enough, as for instance, when we wish to infer a weighted connectivity matrix or parameters like the synaptic delay or a time constant from a biophysical model.

Model structures and/or parameter sets in a certain class can be compared with each other using relative quality measures like the likelihood ratio and other criteria that penalize larger complexity models like the Akaike Information Criterion (AIC) and the Bayesian Information Criterion (BIC) \citep{aho2014model}. 

An alternative strategy is cross validation, in which the inferred model is tested against a separate test data set that was not used for model inference. A standard method in probabilistic model-based methods is to compute a normalized likelihood of the model for a test data set \citep{gerwinn2010bayesian}. 
In addition, comparison of any statistical features, such as average firing rates and spike-train statistics of the sample data produced from the inferred model and the real data is helpful in validating the inferred model \citep{pillow2005prediction}.
When a certain graph-theoretic feature of the network is known, for example, by previous anatomical studies, comparison of such features, such as the node degree distribution, can be a helpful way of validation \citep{bullmore2009complex}.

\section{Challenges}
\label{challenges}

This section summarizes two types of challenges in connectivity inference: biophysical and technical. 
We first describe complexities arising from biophysical properties of neurons and synapses, and then from technical difficulties due to constraints in instrumentation and computation.

\subsection{Biophysical Challenges}
\label{biophysical_challenges}

\begin{description}
  \item [Apparent connectivity:]
As mentioned in the Introduction, functional connectivity (FC) and effective connectivity (EC) may not be due to direct synaptic connectivity. A typical example is a common input that activates two or more neurons simultaneously. In such a case, functional connectivity can be inferred between the recipient neurons even if there is no direct connection between them.
Other cases are a common input with different time delays or an indirect connection through a third neuron, which can activate two neurons in sequence. In this case, effective connectivity can be inferred even if there is no direct synaptic connection between them.

  \item [Directionality:]
Methods based on simple correlation cannot detect the direction of the causal or temporal relationship. Even when the temporal order of activities is considered, if the time resolution of the measurement or analysis is coarse, ordered activation can appear as simultaneous activation, making it difficult to determine directionality.

  \item [Cellular diversity:] 
Many neurons show intrinsic refractoriness such that spike frequency gradually drops even with a constant level of input.
Some neurons also show a burst property such that once excited above a threshold, they keep spiking even without excitatory inputs. 
In such cases, it is not straightforward to discriminate whether a change in the activity of a neuron is due to some network dynamic or the neuron's intrinsic properties 
\citep[chap. 1]{Gerstner2014nd}.
Such refractoriness or burstiness can be categorized as inhibitory or excitatory self-connection, but whether it actually results from self-connecting synapses (autapse) \citep{Bekkers2009cb} must be carefully interpreted.
Even in a local circuit, there are many different types of neurons with a large diversity of biophysical parameters \citep{kandel2000principlesCH4}.
Thus a simple assumption of uniform cellular properties may not be valid.
  \item [Synaptic diversity:]
  Diversity also applies to connectivity because underlying synapses can have complex and diverse characteristics (e.g. excitatory/inhibitory, facilitatory/depressive, delays, etc.) \citep{kandel2000principlesCH8}.
Understanding how a given effective network emerges may require the inference of many additional parameters beyond a simple weight matrix. 

  \item [Non-stationarity:]
Synaptic weights are subject to both short-term (from tens of milliseconds to few minutes) and long-term changes (from minutes to hours) \citep[chap. 19]{Gerstner2014nd}.  
Physiological states of neurons can also drift over time, especially under experimental manipulations, as in slice preparations, with electrodes inserted, or with exposure to light. 
\textit{In vitro} cultured neural networks often manifest switching between states of low and high firing rates.
When using fluorescence imaging, the high firing state can cause erraneous connectivity inference because the low temporal resolution
 makes it difficult to discriminate between direct and indirect effects \citep{stetter2012model}.   
\end{description}

\subsection{Technical Challenges}
\label{technical_challenges}

\begin{description}
  \item [Noise:] 
Every instrument is subject to noise. Electrodes can pick up both biophysical noise (e.g. thermal noise) and anthropogenic noise (e.g. electromagnetic interference from power lines) \citep{van2006signal3noise}.
Optical imaging is susceptible to light scattering artifacts, and motion artifacts in awake subjects with \textit{in vivo} imaging. Even though motion correction programs can track shifts of neurons within an image, movement of cells out of the focal plane is difficult to compensate.

  \item [Time/space resolution:]
Multi-electrode recording can monitor activities of a few hundred neurons with sampling rates on the order of ${10\,kHz}$ \citep{shimono2014functional}.
Fluorescence imaging can work with a trade-off of space and time, ranging from few hundred neurons recorded at ${100\,Hz}$
to a hundred thousand neurons, or $80\,\%$ of a zebra fish brain, at a rate of ${0.8\,Hz}$ \citep{ahrens2013whole}. 
The poor temporal resolution of fluorescence imaging has the undesirable feature of mixing indirect causal effects with real direct connectivity effects. 

  \item [Hidden neurons/External inputs:]
Even though recent advances in two-photon imaging have enabled us to capture thousands of neurons at one time \citep{ahrens2012brain}, it is still difficult to simultaneously record the activity of every single neuron in a circuit.
Neglection of hidden neurons can lead to spurious detections of connectivity between neurons connected via the hidden neurons.
Even though most experiments take into account sensory stimuli or motor actions in the analysis of neural activities, the brain shows varieties of intrinsic dynamics. Unknown external inputs, especially those common to multiple neurons, can also result in a detection of spurious connections.

  \item [Prior knowledge:]
Prior knowledge about the anatomy and physiology of the neural network under investigation can be leveraged to enhance connectivity inference.
How to incorporate prior knowledge about the composition of neuron types in a population \citep{shimono2014functional, sohn2011topological} or connection probabilities between different type of neurons  \citep{white1986structure} to constrain the solution space is an important issue in connectivity inference.
Additional data, such as the local field potential \citep{Destexhe:2013}, can also facilitate understanding of  non-stationary behaviors at the microcircuit level.

  \item  [Scalability:] 
As the number of neurons measured simultaneously grows from hundreds to thousands, the number of potential connections grow from tens of thousands to millions. Therefore, inference methods need to be designed to maximize computational efficiency and to minimize the cost of parallelized implementations \citep[chap. 10]{Gerstner2014nd}.

\end{description}

\section{Model-Free Methods}
\label{modelfreemethods}

We first review \textit{model-free} methods, which do not assume any mechanism to generate the observed data.
These methods tend to be simpler than model-based methods, but they are not able to generate activity data for model validation or prediction. 
We will review model-free methods in two categories. The first is based on descriptive statistics and the second 
on information theoretic concepts. 
Table \cref{tab:model-free1,tab:model-free2,tab:model-free3} summarize the major model-free methods and examples that use those methods.

\newcommand{\chprefix}{1}
\renewcommand\thetable{\chprefix.\arabic{table}}

\newcolumntype{K}[1]{>{\raggedright\arraybackslash}p{#1}}

\begin{table}[p]
  \vspace*{-2.5cm}
  \caption{Summary of model-free connectivity inference methods, Descriptive Statistics (\ref{statisticalmethods}).}
  \label{tab:model-free1}
  %\footnotesize
  \scriptsize
  %\tiny
  %\centering
  \hspace*{-2cm}
  %\begin{tabular}{K{2cm}K{3.5cm}K{3.5cm}K{5cm}K{3cm}}
  \begin{tabular}{K{3cm}K{3cm}K{5cm}K{5cm}}

\toprule[1.0pt]
%Method name (group and section) & Input (most common) & Output & Remarks & References and data source\\
Method & Principle & Examples of application & Features\\

%\midrule[1.0pt]
%\multicolumn{4}{l}{Descriptive Statistics (\ref{statisticalmethods})} \\

\midrule[0.5pt]  
%%%Cross-correlation    
%%%Correlation    
   Correlation (\ref{section:correlation}) & 
   %Spike trains from 2 neurons & 
   Linear relationship &
   \cite{magrans}: simulation data that assumes calcium imaging \newline
   \cite{fujisawa2008behavior}: Multi-electrode recordings from medial prefrontal cortex of rats\newline
   \cite{cohen2011measuring} &
   $\bullet$ Acausal indicator \newline
   $\bullet$ Does not discriminate direct/indirect effects \newline
   $\bullet$ Low computational cost \\

\midrule[0.5pt]  
%%%Cross-correlation    
   Cross-correlation (\ref{section:cc}) & 
   %Spike trains from 2 neurons & 
   Linear relationship with time shift &
   \cite{garofalo2009evaluation}: ROC and PPC curves for CC below JE and TE \newline 
   \cite{ito2011extending}:ROC curve for CC below TE variants curves
  \cite{knox1981detection}\newline
   \cite{garofalo2009evaluation}: simulation data and multi-electrode recordings from in vitro cultured cortical neurons\newline
   \cite{ito2011extending}: simulation data &
   $\bullet$ Causal indicator\newline
   %$\bullet$ Does not discriminate direct/indirect effects\newline
   $\bullet$ Takes into account the spike train history \newline
   $\bullet$ Low computational cost \\  

\midrule[0.5pt]  
%%%Partial-correlation    
   Partial-correlation (\ref{sec:partialcorrelation}) & 
   %Spike trans from all neurons &
   Linear relationship excluding the effect from other neurons &
   \cite{sutera}: simulation data that assumes calcium imaging &
   $\bullet$ Acausal indicator\newline
   $\bullet$ Discriminates direct/indirect effects\newline
   $\bullet$ Solution based on PC won First Neural Connectomics Challenge \\

\bottomrule[1.0pt]
  \end{tabular}
\end{table}

\newcolumntype{K}[1]{>{\raggedright\arraybackslash}p{#1}}

\begin{table}[p]
  \vspace*{-0.5cm}
  \caption{Summary of model-free connectivity inference methods, Information Theoretic (\ref{informationtheoreticmethods}) and Supervised Learning (\ref{section:supervised}, \ref{modelfree_studies3}).}
  \label{tab:model-free23}
  %\footnotesize
  \scriptsize
  %\tiny
  %\centering
  \hspace*{-2cm}
  %\begin{tabular}{K{2cm}K{3.5cm}K{3.5cm}K{5cm}K{3cm}}
  \begin{tabular}{K{3cm}K{3cm}K{5cm}K{5cm}}

\toprule[1.0pt]
%Method name (group and section) & Input (most common) & Output & Remarks & References and data source\\
Method & Principle & Examples of application & Features\\

\midrule[0.5pt] 
%%%Mutual information    
   Mutual information (\ref{sec:mutualinformation}) & 
   %Spike trains from 2 neurons & 
   Statistical dependence &
   \cite{garofalo2009evaluation}: ROC and PPC curves for MI below TE, JE and CC curves  \newline
   \cite{garofalo2009evaluation}: simulation data and multi-electrode recordings from in vitro cultured cortical neurons &
   $\bullet$ Acausal indicator \newline
   $\bullet$ Does not discriminate direct/indirect effects \newline
   $\bullet$ Does not discriminate between excitatory and inhibitory connections \\
   
\midrule[0.5pt] 
%%%Joint entropy    
   Joint entropy (\ref{section:jointentropy}) & 
   %Spike trains from 2 neurons & 
   Entropy of cross-inter-spike-intervals &
   \cite{garofalo2009evaluation}: ROC and PPC curves for JE below TE and above MI and CC curves \newline
   \cite{garofalo2009evaluation}: simulation data and multi-electrode recordings from in vitro cultured cortical neurons &
   $\bullet$ Acausal indicator\newline
   $\bullet$ Does not discriminate direct/indirect effects\newline
   $\bullet$ Does not discriminate between excitatory and inhibitory connections \\

\midrule[0.5pt] 
%%%Transfer entropy    
   Transfer entropy (\ref{section:transferentropy}) & 
   %Spike trains from 2 neurons & 
   Gain in past-future mutual information when the activity of another neuron is considered &
   \cite{garofalo2009evaluation}: ROC and PPC curves for TE above the rest (JE, MI and CC curves) \newline
   \cite{schreiber2000measuring}\newline  
   \cite{garofalo2009evaluation}: simulation data and multi-electrode recordings from in vitro cultured cortical neurons\newline
   \cite{shimono2014functional}: multi-electrode recordings from from slice cultures of rodent somatosensory cortex&
   $\bullet$ Causal indicator\newline
   $\bullet$ Does not discriminate direct/indirect effects\newline
   $\bullet$ Does not discriminate between excitatory and inhibitory connections. However, it is possible to perform a subsequent analysis of the correlation or conditional distribution to infer if it is excitatory or inhibitory\newline
   $\bullet$ Takes into account the spike train history \\

\midrule[0.5pt]  
%%%Delayed transfer entropy    
   Delayed transfer entropy (\ref{section:dte}) & 
   %Spike trains from 2 neurons & 
   Transfer entropy with delayed source signal &
   \cite{ito2011extending}: ROC curve and TPR at constant FPR better than TE and below HOTE \newline
   \cite{ito2011extending}: simulation data &
   $\bullet$ Same as transfer entropy \\  
 
\midrule[0.5pt]  
%%%High order transfer entropy    
   High order transfer entropy (\ref{section:hote}) & 
   %Spike trains from 2 neurons & 
   Transfer entropy with multiple time steps of activities &
   \cite{ito2011extending}: ROC curve and TPR at constant FPR better than TE and DTE \newline
   \cite{ito2011extending}: simulation data &
   $\bullet$ Same as transfer entropy \\  

\midrule[0.5pt]    
%%%Generalized transfer entropy    
   Generalized transfer entropy (\ref{section:gte}) & 
   %Spike trains from 2 neurons and time series of the overall network activity & 
   Transfer entropy with present source activity, after removing high activity periods &
   \cite{stetter2012model, orlandi2014transfer}: ROC curve for GTE above TE curve \newline
   \cite{stetter2012model, orlandi2014transfer}: simulation data that assumes calcium imaging and data recorded using calcium imaging from in vitro networks derived from cortical neurons  &  
   $\bullet$ The use of information from the same time bin for both neurons enhances reconstruction performance when the data source has a low sampling rate (e.g. calcium imaging) \newline
   $\bullet$ It copes well with synchronized bursting episodes \\
 
\midrule[0.5pt] 
 %%%Convolutional neural network and other supervised methods    
   Convolutional neural network (\ref{modelfree_studies3}) &
   %$\bullet$ Training time: activity from 2 neurons (e.g. spike trains, fluorescence time series, connectivity measures) and connectivity labels   \newline
   %$\bullet$Inference time: activity from 2 neurons (e.g. spike trains, fluorescence time series, connectivity measures) & 
   Probability of connection generalized from training data &
   \cite{lukasz,veeriah2015deep}: simulation data that assumes calcium imaging &
   $\bullet$ Assumes the ground truth connectivity labels of a training data set that has similar properties as the network that we want to reconstruct \\

\bottomrule[1.0pt]
  \end{tabular}
\end{table}

We denote the neural activity data set as
$$D=\left\{ x_{i}(t)| i=1,\dots, P; t=1,\dots, T\right\},$$ where $x_{i}(t)$ is the activity of neuron $i$ at the $t^{th}$ time point.
$x_{i}(t)$ may be continuous, for instance, when the values are raw data obtained from multiple electrodes or fluorescence imaging, or it may be binary when data are transformed into a spike train by a spike-sorting algorithm.

\subsection{Descriptive Statistics}
\label{statisticalmethods}

This class of methods utilize statistical measures to capture the degree of connection between neurons from a sample of activities from a neural population.

\subsubsection{Correlation}  
\label{section:correlation}

Correlation indicates the strength of the linear relationship between two random variables that represent two neurons.
The most commonly used measure of correlation between activities $x_i$ and $x_j$ of two neurons $i$ and $j$ is the Pearson correlation coefficient defined as:
\begin{equation}
\label{eqn: correlation}
 \rho_{ij}= 
 \frac{\Sigma_{ij}} {\sqrt{\Sigma_{ii} \Sigma_{jj}}},
\end{equation}
where $\Sigma_{ij} = \frac{1}{T} \sum_{t=1}^{T} (x_{i}(t)-\mu_{i})(x_{j}(t)- \mu_{j})$ is the covariance and $\mu_{i} = \frac{1}{T} \sum_{t=1}^{T} x_{i}(t)$ is the mean activity.

When we use correlation $\rho_{ij}$ to perform network inference of neuronal circuits, we are measuring the rate of co-occurring spikes of neurons $i$ and $j$~\citep{cohen2011measuring} and this rate is interpreted as the functional connectivity strength. 
While Pearson correlation is computationally least costly, it has a number of drawbacks. It is not able to indicate the causal direction and it is not able to distinguish direct connections from indirect ones. It is also not suited to deal with external inputs. 
Despite such limitations, a winning solution of the First Neural Connectomics Challenge used correlation as a key component in a more complex method (see section \ref{modelfree_studies2}).

\subsubsection{Cross correlation}
\label{section:cc}

Cross correlation (CC) indicates the strength of the delayed linear relationship between two neurons $i$ and $j$. \cite{knox1981detection} defines CC as "the probability of observing a spike in one train $x_j$ at time $t+\tau$, given that there was a spike in a second train $x_i$ at time $t$". \cite{ito2011extending} remark that, despite the extensive literature, there is not a standard definition and they discuss the two most popular ones:
\begin{equation}
 \label{eqn: crosscorrelation4}
\rho_{i \rightarrow j}(\tau) =\frac{1}{T-1} \sum_{t=1}^{T} \frac{(x_{i}(t)-\mu_{i})(x_{j}(t+\tau)- \mu_{j})}{\sigma_i\sigma_j}
\end{equation}
where the parameter $\tau$ defines the delay of neuron $j$ with respect to neuron $i$, and $\mu$ and $\sigma$ are the sample average and standard deviation, respectively.

The second CC definition uses the total number of spikes instead of the averages and standard deviations:
\begin{equation}
 \label{eqn: crosscorrelation2}
  \rho_{i \rightarrow j}(\tau) = \sum_{t=1}^{T} \frac{x_{i}(t)x_{j}(t+\tau)} {\sqrt{n_{i} n_{j}}},
\end{equation}
where $n_{i}$ and $n_{j}$ are the total number of spikes from neuron $i$ and $j$ respectively. Both cross-correlation definitions tend to be equivalent when $\mu_{i}$ and $\mu_{j}$ approaches 0. 

CC is a causal indicator which is able to indicate the direction of the connection. The inference performance of the connection direction is dependent on the instrumentation sampling rate and the choice of the time delay $\tau$.
Despite the added capability of being able to detect the direction of connectivity, CC has the same limitations as correlation in dealing with indirect connections and external inputs.

We should also note that the optimal parameter $\tau$ to detect a connection can be different for each connection. A measure to address this issue is the Coincidence Index (CI), which combines several CCs computed at different $\tau$s:
 \begin{equation}
  \label{eqn: coincidenceindex}
   CI_{i \rightarrow j} = \frac{\sum_{\tau=0}^{r}\rho_{i \rightarrow j}(\tau)}
   {\sum_{\tau=0}^{T}\rho_{i \rightarrow j}(\tau)},
\end{equation} 
where $r$ specifies the interval of cross-correlation delays, called the coincidence interval.
% and $T$ is the full window size of cross-correlation delays. 
% Isn't $T$ the length of the data??
A large CI indicates a larger reproducibility of correlated spike timing \citep{tateno1999activity,chiappalone2007network,shimono2014functional}. 

\cite{garofalo2009evaluation} evaluated the network inference performance of several model-free methods
% (cross-correlation, mutual information, transfer entropy and joint entropy) 
using data from simulated neuronal networks with only excitatory synapses and including both excitatory and inhibitory connections. 
According to both ROC and PPC criteria, CC demonstrated a performance just below transfer entropy (TE, see section \ref{section:transferentropy}) in the fully excitatory setting, and below TE and joint entropy (JE, see section \ref{section:jointentropy}) when inhibitory connections were also included.
 \cite{ito2011extending} evaluated the inference performance of the two CC variants discussed in this section and several transfer entropy variants (see sections \ref{section:dte}, \ref{section:hote} and \ref{section:gte}) using data from simulated neuronal networks with both excitatory and inhibitory connections. According to the ROC criterion, both CC variants showed a performance inferior to all variants of transfer entropy. 

\subsubsection{Partial correlation}
\label{sec:partialcorrelation}
  
Let $R = [R_{ij}]$ be an inverse of a covariance matrix $\Sigma$ of which the $(i,j)^{th}$ component
is $\Sigma_{ij}$.
The definition of partial correlation (PC) between activities of neurons $i$ and $j$ is:
\begin{equation}
 \label{eqn: partialcorrelation}
  PC_{ij} = - \frac{R_{ij}}{\sqrt{R_{ii} R_{jj}}}.
\end{equation}  
The most salient difference between PC and other methods described in this section is that PC takes into account the activities of all neurons in the population to compute a connectivity indicator between each neuron pair. 
An important property of PC is that, assuming that $x(t)=(x_{1}(t),\dots, x_{P}(t))$ is
normally distributed, then $PC_{ij}$ is $0$ if and only if neurons $i$ and $j$ are independent given all the rest. Therefore, PC can be used to distinguish a direct effect from an indirect one. 
Despite this multivariate feature, PC shares the same limitations as correlation and CC in dealing with external inputs, with the additional computational complexity required to invert the covariance matrix. 
The first prize solution of the First Neural Connectomics Challenge, discussed in section \ref{modelfree_studies1}, is a good example of how to use PC for network inference \citep{sutera}. 

\subsection{Information Theoretic Methods}
\label{informationtheoreticmethods}

Information theory is a mathematical discipline initiated by Shannon to characterize the limits of information management, including transmission, compression and processing \citep{shannon1948mathematical}. This section presents the application of several information theory measures to the inference of neural microcircuits.

\subsubsection{Mutual information}
\label{sec:mutualinformation}

Mutual information (MI) is a measure of the statistical dependence between stochastic variables.
MI of activities $x_i$ and $x_j$ of two neurons is mathematically defined as:
\begin{equation}
  \label{eqn: mutualinformation2}
	 MI_{i,j} =  \sum_{i,j} P(x_i,x_j) \log {\frac {P(x_i,x_j)}{P(x_i)P(x_j)}}.
\end{equation} 
This indicator is symmetric; therefore it is unable to identify the direction of the connection. It cannot discriminate direct effects from indirect effects.  
One way to overcame the directionality limitation is to introduce a delay as in CC, and to consider multiple delays as in CI \citep{garofalo2009evaluation}.

In a comparative study by \cite{garofalo2009evaluation} using data from simulated neuronal networks, MI delivered the performance below TE, JE and CC, according to both ROC and PPC criteria.  
 
\subsubsection{Join entropy}
\label{section:jointentropy}

Joint entropy (JE) is a bi-variate causal measure between the activity of two neurons, testing whether neuron $i$ is a cause of the activity in neuron $j$. 
For each reference spike in $x_i$, the closest future spike on $x_j$ is identified and the cross-inter-spike-interval is computed as $cISI=t_{x_i}-t_{x_j}$, where $t{x_i}$ and $t_{x_j}$ are the time of successive spikes of neuron $i$ and $j$, respectively. Then JE is defined as the entropy of the cISI distribution:
\begin{equation}
  \label{eqn: joinentropy}
  JE_{i \rightarrow j} = -\sum_{k} P(cISI_{k}) \log_2 P(cISI_{k}),
\end{equation}
where $k$ indexes the level of cISI. 
If neurons $i$ and $j$ are strongly connected, then JE is expected to be close to 0 as the cISI distribution becomes sharp.

In a comparison by \cite{garofalo2009evaluation}, JE showed the worst performance below TE, CC and MI in a fully excitatory setting, but the second best performance below TE when inhibitory connections were also included .

\subsubsection{Transfer entropy}  
\label{section:transferentropy} 

 Transfer entropy (TE) is a causal indicator defined between the activity of two neurons $i$ and $j$.
% (note that we are using a slightly different notation to simplify the equations of transfer entropy and its variants: $x$ and $y$ are the activity vectors of neuron $X$ and $Y$ respectively instead using $x_i$ and $x_j$ to denote the activity of neurons $i$ and $j$ respectively). 
The definition of TE was originally formulated as the Kullback-Leibler divergence between the distributions of the neural activity of target neuron $j$ conditioned by its previous activities alone versus conditioned also by previous activities of source neuron $i$ \citep{schreiber2000measuring}:  
\begin{eqnarray}
  \label{eqn: transferentropy}
%  TE_{Y \rightarrow X} = \sum_{x_{t+1},x_{t}^k,y_{t}^l} P(x_{t+1},x_{t}^k,y_{t}^l) \log_2 \frac {P(x_{t+1}|x_{t}^k,y_{t}^l)}{P(x_{t+1}|x_{t}^k)} 
  TE_{i \rightarrow j} = &\sum&
  %_{x_j(t),x_j(t-1),...,x_j(t-k),x_i(t-1),...,x_i(t-l)} 
    P(x_j(t),x_j(t-1),...,x_j(t-k),x_i(t-1),...,x_i(t-l)) \nonumber \\
    && \log_2 \frac {P(x_j(t)|x_j(t-1),...,x_j(t-k),x_i(t-1),...,x_i(t-l))}{P(x_j(t)|x_j(t-1),...,x_j(t-k))}.
\end{eqnarray}  
%where $x_{t}^k$ is the spike train generated by neuron $X$ from time $t-k$ to time $t$, $y_{t}^l$ is the spike train generated by neuron $Y$ from time $t-l$ to time $t$. 
TE can also be defined as the conditional mutual information between the future activity of target neuron $j$ and past activity of source neuron $i$ conditioned by past activity of target neuron $j$, which can also be expressed as the amount of uncertainty reduced in the future activity of neuron $j$ knowing the past activity of $i$ given past activity of $j$:

\begin{eqnarray}
  \label{eqn: transferentropy2}
%  TE_{Y \rightarrow X} = MI(x_{t+1},(y_{t}^l,x_{t}^k))-MI(x_{t+1},x_t^k) 
  TE_{i \rightarrow j} &=& MI(x_j(t),(x_i(t-1),...,x_i(t-l)) | (x_j(t-1),...,x_j(t-k)))\nonumber \\
  							&&=H(x_j(t)|(x_j(t-1),...,x_j(t-k))) \nonumber \\ 
  							&&-H(x_j(t)|(x_j(t-1),...,x_j(t-k),x_i(t-1),...,x_i(t-l))) 
\end{eqnarray}  
From this second definition, we can see that TE is a positive measure that takes a low value when the past activity of neuron $i$ does not convey information about the future activity of neuron $j$.
%The larger the influence of past values of $Y$ to predict the future value of $X$, the more divergent these two distributions will be.  
% Interestingly, TE and cross-correlation are the two only model-free method that takes into account the spike train history. 

TE is equivalent to Granger causality for Gaussian variables \citep{barnett2009granger}. A nice feature of TE is that it can detect non-linear relationships \citep{stetter2012model}. 
Although TE does not identify whether the connection is excitatory or inhibitory, the polarity of connection can be tested separately by recording (real data) or simulating our network with the inhibitory connections pharmacologically blocked \citep{stetter2012model,orlandi2014transfer}. 
The performance of TE in discriminating the causal direction depends whether the sampling rate is faster than the network dynamics \citep{fujisawa2008behavior, shimono2014functional}). 
%(e.g. sampling rate from a microelectrode array system $\sim 20\,kHz$ 
%\cite{ito2011extending} show that the reconstruction performance of inhibitory connections is strongly dependent on the postsynaptic neuron to have a high firing rate.

In practice, the parameters $k$ and $l$ in Eq.~\eqref{eqn: transferentropy} are set to $1$ (e.g.,  \cite{garofalo2009evaluation}), so that we only have to consider a small number of patterns (i.e. $2^3=8$ for binary data). While increasing $k$ and $l$ can allow detection of delayed effects of a connection, it requires larger amount of data to have reliable estimates of the probability distribution to compute the entropies.

%An estimator of TE for discrete variables is based on growing a histogram for each probability estimate \citep{lizier2014jidt} has a computational complexity of $O(T)$ where $T$ is the number of time steps.  

Before the First Neural Connectomics Challenge, TE was arguably the most successful model-free method \citep{garofalo2009evaluation,stetter2012model}. This success encouraged a number of variants as follows \citep{ito2011extending,stetter2012model, orlandi2014transfer}.
  
%\subsubsection{Variants of transfer entropy}
%\label{section:vte}
 
\subsubsection{Delayed transfer entropy (DTE)} 
\label{section:dte}
A limitation of TE using $k=l=1$ is that it assumes a constant one-time bin delay between the action potential in neuron $i$ and the post-synaptic action potential of neuron $j$, which is not a realistic assumption. A suboptimal way to deal with this problem is to create longer time bins at the cost of losing detailed temporal information. Instead, delayed TE (DTE) was proposed to measure TE at a user-defined time delay $d$ \citep{ito2011extending}. Compared to the original definition, instead of $x_i(t-1),...,x_i(t-l)$ in Eq.~\eqref{eqn: transferentropy}, they suggest using a delayed signal $x_i(t-d),...,x_i(t-d-l+1)$. \cite{shimono2014functional} improves reconstruction performance by identifying the delay parameter for each individual connection using the coincidence index (Eq.~\eqref{eqn: coincidenceindex}). 
  
\subsubsection{High order transfer entropy (HOTE)} 
\label{section:hote}
%When using binary spike trains, usually (e.g. \citep{garofalo2009evaluation}) the parameters $k$ and $l$ in Eq.~\eqref{eqn: transferentropy} are set to $1$, so we only have to consider a low number of patterns (i.e. $2^3=8$). 
\cite{ito2011extending} considered how to increase $k$ and $l$ to take into account longer spike train history, while avoiding negative effects of the increase in the possible patterns ($2^{k+l+1}$).
%This has the added cost of having a larger number of possible patterns ($2^{k+l+1}$) and therefore a reduced number of instance of each pattern given the same amount of data.  
The additional parameters $d$ in DTE and $k$ and $l$ in HOTE gives multiple measures for each neuron pair.
%\cite{ito2011extending} use the peak value and the coincidence index (see section 4.1.2) to be able to compare measurements across neuron pairs. 
They evaluated the reconstruction performance by the peak value and the coincidence index (Eq.~\eqref{eqn: coincidenceindex}) of DTE and HOTE, using a simulated network with both excitatory and inhibitory connections. 
%According to the ROC curve and the TPR at a fixed FPR=0.01, 
HOTE$_{CI}$ and DTE$_{CI}$ had better performance than HOTE$_{pk}$ and DTE$_{pk}$, which in turn were better than TE with $k=l=1$. 
%HOTE$_{CI}$ and HOTE$_{pk}$ delivered the best reconstruction performance of inhibitory connections followed by DTE$_{CI}$ and DTE$_{pk}$.

%Examples: \ref{tab:model22}

\subsubsection{Generalized transfer entropy (GTE)} 
\label{section:gte}
As described in section \ref{challenges}, periods of synchronized bursting convey very low connectivity information due to the simultaneous spike of a large percentage of neurons. This phenomenon is especially critical with calcium imaging recordings because of their limited temporal resolution. 
Generalized TE \citep{stetter2012model} was proposed to alleviate this problem
%\begin{multline}
%  \label{eqn: generaltransferentropy}
%  GTE(\tilde{g})_{Y \rightarrow X} = \\   \sum_{x_{t+1},x_{t}^k,y_{t+1}^l} P(x_{t+1},x_{t}^k,y_{t+1}^l|g_{t+1}<\tilde{g}) \log_2 \frac {P(x_{t+1}|x_{t}^k,y_{t+1}^l,g_{t+1}<\tilde{g})}{P(x_{t+1}|x_{t}^k,g_{t+1}<\tilde{g})} 
% \end{multline} 
%where $g_{t+1}$ is the overall network activity at time $t+1$ and $\tilde{g}$ is a user-defined threshold. 
by two modifications to equation~\eqref{eqn: transferentropy}: 
i) to compensate for the slow sampling rate in fluorescence imaging ($\geq 10ms$),
% which is longer than the lower range of synaptic delays (e.g. conduction delays in the mammalian nervous system range from $<$ 100 $\mu$s to $>$ 100 ms \citep{Swadlow:2012}). 
GTE uses presynaptic activity from the same time bin $x_i(t)$ instead of the previous one $x_i(t-1)$; 
ii) to discard the poor connectivity information conveyed by synchronized bursting periods, GTE restricts the computation of TE to those time bins with an overall network activity below a given user-defined threshold. 

Just like TE, GTE is not able to differentiate between excitatory and inhibitory connections. 
The application of GTE to simulated neural networks with excitatory connections shows an improved performance with respect to TE, and modified versions of CC and MI implementing the same generalization \citep{stetter2012model}. \cite{orlandi2014transfer} showed that GTE is able to reconstruct inhibitory connections; however, its reconstructions performance is below its reconstruction performance of excitatory connections. 
%\end{enumerate}
  
\subsubsection{Information gain}
Information gain (IG) is a less known causal indicator between the activity of two neurons $i$ and $j$. Its definition is:
\begin{equation}
  \label{eqn: informationgain}
%  IG_{i \rightarrow j} = IG(x_j|x_i) = H(x_j)-H(x_j|x_i) \approx _{MEA} H(x_j(t+1))-H(x_j(t+1)|x_i(t))  \\
  IG_{i \rightarrow j} = H(x_j(t)) - H(x_j(t)|x_i(t-1)),
 \end{equation} 
where $H$ is the entropy, $x_i$ is the activity vector of the source neuron and $x_j$ is the activity vector of the target neuron. 
%In the case of MEA, due to its high sampling rate, samples from the target neurons are taken ahead in time with respect to the samples of the source neuron. 
In the case of calcium imaging recordings, it may be convenient to employ the same strategy as with GTE by using the sample from the same time bin $x_i(t)$ instead of the previous one $x_i(t-1)$. 
This connectivity indicator was used as one of multiple indicators combined by one of the top solutions of the First Neural Connectomics Challenge \citep{Jozefowicz}.

\subsection{Supervised Learning Approach}
\label{section:supervised}

Motivated by recent successes of deep neural networks in challenging patter classification problems, a new approache has been proposed to apply convolutional neural networks (CNNs) for prediction of the existence of synaptic connectivity \citep{lukasz,veeriah2015deep}.
In this approach, the time series data of spike trains or fluorescent signals of a pair or more of neurons are given as the input and a CNN is trained to classify if there is a connection between the neurons.

A natural limitation of this approach is that training requires sufficient amount of ground truth data about the existence of connectivity.
Although this approach has been successful with simulated data \citep{lukasz,veeriah2015deep}, it is hard to apply them directly to real data, for which experimental identification of synaptic connectivity is highly costly.  

\section{Model-Based Methods}
\label{modelbasedmethods}

Now we review {\it model-based} methods, in which the connectivity is estimated by explicitly modeling the data generation process.
The basic paradigm of
the model-based method is:
1) to assume a generative model that generates neural data, and
2) to determine the model parameters to fit observed data.
Table \ref{tab:model-based} summarizes the major model-based methods and examples that used those methods.
%%%
%%%
%%% Added by J.Y. on 08 Nov. (Begin here)
%%%
%%%
We should keep in mind here that any model for this approach aims to explain the observed data, but does not consider all physiological properties of the real neural networks: The large number of factors affects structure and function of the real neural systems, and it is infeasible for all of them to be considered in a single model (See more details about similarities and gaps between the model and biological reality in~\citep{Mohamad_Ask_2016}). Due to the limitation, the applicability of the model-based methods is shown only empirically using simulation studies, but there is no theoretical guarantee that the estimation result should be consistent with the underlying connectivity. However, through reviewing the model-based methods here, we will see approximation techniques to fill the gaps with a feasible number of parameters as well as practical solutions for the connectivity estimation.

%%%
%%%
%%% Added by J.Y. on 08 Nov. (End here)
%%%
%%%
\renewcommand{\chprefix}{2}
\renewcommand\thetable{\chprefix}

%\newcolumntype{K}[1]{>{\raggedright\arraybackslash}p{#1}}

\begin{table}[p]
  \caption{Summary of model-based connectivity inference methods.}
  \label{tab:model-based}
  %\footnotesize
  \scriptsize
  %\tiny
  \centering
  \hspace*{-2cm}
  %\begin{tabular}{K{2cm}K{3.5cm}K{3.5cm}K{5cm}K{3cm}}
  \begin{tabular}{K{3cm}K{3cm}K{5cm}K{5cm}}

\toprule[1.0pt] 
%Model (section) & Input attributes (most common) & Output & Remarks & References and data source\\
   Method & Principle & References \& demonstrations
	   {\tiny (The asterisk ($^{\ast}$) shows inclusion of applications to real data)}& Features\\

\midrule[1.0pt]  
   %%% AR model
   Autoregressive models (\ref{section:ARmodel}) &
   % multivariate continuous time-series
   Directed linear interaction &
    $^{\ast}$\cite{Harrison2003ni}: human fMRI data \newline
    %\cite{valdes2005estimating}: human fMRI data \newline
    $^{\ast}$\cite{Franaszczuk1985bc}: multichannel EEG time series \newline
	\hspace*{0.5em}\cite{Smith2011ieeeembc}: simulation data that assumes calcium imaging &
	$\bullet$ analytical solutions are available \newline
	$\bullet$ not suitable for spike-train data \\

\midrule[0.5pt]
    %%% GLM model
    Generalized linear model (\ref{section:glm}) &
    % multineuronal spike trains
	Spike probability based on linear summation of inputs &
	$^{\ast}$\cite{Song2013jcn}: simulation data and multi-electrode recordings from hippocampal CA3 and CA1 regions of rats \newline
	$^{\ast}$\cite{gerwinn2010bayesian}: simulation data and multi-electrode recordings from salamander retinal ganglion cells &
	$\bullet$ easy but iterative optimization is required \\%\newline
	%$\bullet$ the number of parameters can be reduced by appropiate basis functions\\
	
\midrule[0.5pt]  
   %%% stochastic LIF
   Stochastic leaky integrate-and-fire model (\ref{section:SLIF}) & 
   % multineuronal spike trains & 
   Stochastic spike with state resetting &
   \hspace*{0.5em}\cite{Paninski2004nc}: simulation data\newline
   \hspace*{0.5em}\cite{Koyama2010jcn}: simulation data\newline
   \hspace*{0.5em}\cite{Isomura2015nc}: simulation data &
   $\bullet$ model for a continuous time domain\newline
   $\bullet$ a special case of generalized linear models \\

\midrule[0.5pt] 
   %%% network likelihood model
   Network likelihood model (\ref{section:NLM}) &
	% multineuronal spike trains
   Continuous-time model with Poisson spikes &
   $^{\ast}$\cite{Okatan2005nc}: simulation data and multi-electrode recordings from hippocampal CA1 region of a rat\newline
   $^{\ast}$\cite{Kim2011ploscb}: simulation data and multi-electrode recordings from the primary motor cortex (MI) of a cat\newline
   $^{\ast}$\cite{Stevenson2009ieee}: simulation data and multi-electrode recordings from the primary motor cortex (MI) and dorsal premotor cortex (PMd) of a monkey &
   $\bullet$ a special case of generalized linear models\newline
   $\bullet$ instantaneous firing rates are directly considered \\

\midrule[0.5pt] 
   %%%Calcium fluorescence model
   Calcium fluorescence model (\ref{section:Calcium}) &
   % multineuronal spike trains
    &
   \hspace*{0.5em}\cite{Mishchenko2011aas}: simulation data \newline &
   \\

\midrule[0.5pt] 
   %%%Hawkes process model
   Hawkes process (\ref{sec:hawkes}) &
   % multineuronal spike trains
   GLM with Poisson observation &
   \hspace*{0.5em}\cite{linderman2014discovering} \newline
   \hspace*{0.5em}\cite{linderman2015scalable}: simulation data that assumes calcium imaging &
   $\bullet$ The network consists of purely excitatory neurons \\

\midrule[0.5pt]  
   %%% Dynamic Bayesian Network
   Dynamic Bayesian network (\ref{section:dbn}) &
   % multivariate continuous time-series or multineuronal spike trains
   Directed acyclic graph &
	\hspace*{0.5em}\cite{eldawlatly2010use}\newline
	$^{\ast}$\cite{patnaik2011discovering}: simulation data and multi-electrode recordings from dissociated cortical cultures of rats &
	       $\bullet$ Simulated annealing for optimizing both network structures and parameters
	       \newline
	       $\bullet$ The computational efficiency is degraded in case of cyclic graphs
	       \newline
	       $\bullet$ Due to the nature of Bayesian approach, the result depends on hypothetical choices of the prior distribution.
	       \\

\midrule[0.5pt] 
   %%%Maximum entropy model
   Maximum entropy model (\ref{section:maxentropy}) &
   % multineuronal spike trains
   Maximum entropy under constraint by an energy function &
   $^{\ast}$\cite{tkacik2014thermodynamics}: multi-electrode recordings from retinal ganglion cells\newline	&   
   %\cite{yeh2010maximum}\newline
   %\cite{roudi2015multi} &
   $\bullet$ Inspired by Ising models\newline
   $\bullet$ Up to the second-order statistics are considered\newline
   $\bullet$ Heavy computational complexity for parameter estimation \\
  
\bottomrule[1.0pt]
  \end{tabular}
\end{table}

In the following, $x(t) = (x_{1}(t), \dots, x_{P}(t))$ represents a set of $P$ signals
observed at the $t^{th}$ time point in a discrete time domain,
where $x_{i}(t)$ denotes the $i^{th}$
neuron's activity at that time. $x_{i}(t)$ may be continuous when the
measurement is raw data collected from calcium imaging and
multiple-electrode recording, or binary when the data is transformed into a
spike train by a spike-sorting algorithm.
We denote a generative model $p_{\theta}(x)$, where $\theta$ is a parameter vector, including the connection weights.

\subsection{Generative Models}

%The common feature of model-based methods is that $x_{i}(t)$ is assumed
%to be a realization sampled from a conditional probability distribution
%$p_{\theta}(\cdot|H(t))$ where $H(t) \equiv \left\{ x(s)|s=1,\cdots,
%t-1\right\}$ is a previous history of the signals and the subscript
%$\theta$ denotes a set of parameters to specify the probability
%distribution.

\subsubsection{Autoregressive model}
\label{section:ARmodel}

A basic example of a generative model is the {\it
autoregressive (AR) model} mathematically expressed as
\begin{eqnarray}
 x_{i}(t) &=& A_{i0}
  + \sum_{j=1}^{P} \sum_{k=1}^{K} A_{ij}(k) x_{j}(t-k) 
  + \epsilon_{i}(t),
  \quad i=1,\cdots,P.
  \nonumber \\
 \epsilon_{i}(t) &\sim& \mathcal{N}(\cdot|0,\sigma_{i}^{2})
  \label{ARmodel}
\end{eqnarray} 
where $K$ is the degree of the model,
$A_{ij}(k)$ is a parameter called an AR coefficient,
$A_{i0}$ is a bias term, and
$\mathcal{N}(\cdot|\mu,\sigma^{2})$ denotes the Gaussian distribution
with mean $\mu$ and standard deviation $\sigma$ (For simplicity,
we assume that $\sigma$ is known in this example).

By integrating two equations in Eq.~\eqref{ARmodel},
$x_{i}(t)$ can be regarded as
a sample according to the following conditional distribution:
\begin{eqnarray}
 x_{i}(t) &\sim& %p_{\theta}(\cdot|H(t)) =
  \mathcal{N}
  \left( \mu_{i}(t) ,\sigma_{i}^{2} \right),
  \\
  \mu_{i}(t) &=&
  A_{i0} + \sum_{j=1}^{P} \sum_{k=1}^{K} A_{ij}(k) x_{j}(t-k)
\end{eqnarray}
where $\theta \equiv \left\{ A_{i0} | i=1,\dots,P \right\} \cup
\{ A_{ij}(k) | i,j=1,\dots, P; k = 1, \dots, K\}$ is a set of parameters
in this case.

%The parameter $\theta$ is determined so that the deviation between the
%model and the observation data is minimized in some sense.
%The simplest approach is {\itshape the least square
%method}, in which $\theta$ is determined to minimize
%{\itshape the sum of squared residuals} defined as
%\begin{equation}
% J(\theta|D) = \sum_{i=1}^{P} \sum_{t=1}^{T} \epsilon_{i}^{2}(t)
%  = \sum_{i=1}^{P} \sum_{t=K+1}^{T} 
%  \left[
%   x_{i}(t) - A_{i0} -
%   \sum_{j=1}^{P} \sum_{k=1}^{K} A_{ij}(k) x_{j}(t-k)
%  \right]^{2},
%\end{equation}
%when $D \equiv
%\left\{ x_{i}(t) | i=1,\dots, P; t=1,\dots, T\right\}$ is given as
%the observation data set.
%From the point of view of statistics, the determination is the same as
%the maximum likelihood method.
 
After determining the parameter $\theta$, 
%we may infer the interaction between any pair of
%the $i^{th}$ and the $j^{th}$ signals as follows.
if $A_{ij}(k)$ for all $k$ is exactly or very close to zero,
%the $j^{th}$ signal $x_{j}$ has contributes little
%to the $i^{th}$ signal $x_{i}$, 
it implies that there is no direct interaction
from the $j^{th}$ neuron the $i^{th}$ neuron.
In contrast, if $A_{ij}(k)$ for some $k$
deviates enough from zero, the $j^{th}$ neuron $x_{j}$ can directly
affect the $i^{th}$ neuron $x_{i}$with a $k$ time-step delay.

Even though neuron's dynamics is usually nonlinear, for the virtue of simplicity, this method has been used
in a wide range of neural data analysis, in addition to calcium
imaging and multiple electrode recording \citep{Harrison2003ni,valdes2005estimating,Franaszczuk1985bc,Smith2011ieeeembc}.

\subsubsection{Generalized linear model}
\label{section:glm}

When $x_{i}(t)$ is a binary variable indicating whether a spike is
generated at the $t^{th}$ time step,
the generalized linear models (GLM)~\citep{Song2013jcn} provides a
tractable extension of the AR models. A GLM describes spike
generation as a point process as
\begin{subequations} %10:19equations
\label{101719_19Aug16} 
 \begin{align}
  x_{i}(t) &\sim \text{Ber}(\cdot | \rho_{i}(t))
  \label{091137_1Oct15} \\
  \rho_{i}(t) &\equiv \phi
  \left(
  A_{i0} + \sum_{j=1}^{P} \sum_{k=1}^{K} A_{ij}(k) x_{j}(t-k)
  \right),
  \label{091141_1Oct15}
 \end{align}
\end{subequations}
where $\text{Ber}(\cdot|p)$ denotes the Bernoulli distribution with
an event probability of $p$.
$\phi(\cdot)$ is a so-called an inverse link function, for which
the exponential function $\phi(x) = \exp (x)$ or 
the sigmoid function $\phi(x) = (1 + \exp (-x))^{-1}$) is often used.

\subsubsection{Stochastic leaky integrate-and-fire model}
\label{section:SLIF}

%Although we have seen GLMs as an extension of AR models so far,
%GLMs are closely related to 
A stochastic leaky integrate-and-fire (LIF)
model~\citep{Paninski2004nc,Koyama2010jcn,Isomura2015nc},
is one of the most widely used model for analyzing the behavior of
spiking neural networks \citep{Gerstner2014nd}.
The model assumes that the subthreshold membrane potential of the
$i^{th}$ neuron, denoted by $V_{i}$, evolves according to the following
stochastic differential equation:
\begin{equation}
 d V_{i}(t) = \left( -g_{i} V_{i}(t) + \sum_{j=1}^{P} I_{ij}(t) \right)dt
  + \sigma_{i} dW_{i}(t),
  \label{134035_28Oct15}
\end{equation}
where $g_{i}$ is the membrane leak conductance and
$dW_{i}(t)$ is an increment of a Wiener process.
$I_{ij}$ is an influence from the $j^{th}$ neuron to the $i^{th}$ neuron,
which is often assumed to be
\begin{equation}
 I_{ij}(t) = \sum_{\{ f:t_{j}^{(f)}<t \}}\kappa_{ij}(t - t_{i}^{(f)}),
\end{equation}
where $\kappa_{ij}(s)$ represents the effect of the the $j^{th}$ neuron to the $i^{th}$ neuron after a time delay $s$ and $t_{j}^{(f)}$ is the $f^{th}$ spike of the $j^{th}$ neuron.
%$N_{j}$ is the total number of the $i^{th}$ neuron's spikes observed before time $t$.

In the hard-threshold version~\citep{Paninski2004nc},
whenever $V_{i}(t)$ goes over a threshold $V_{th}$,
the neuron generates a spike.
In the soft-threshold case~\citep{Koyama2010jcn,Isomura2015nc},
the probability for the neuron to generates a spike in a small
time interval $dt$ is given by
\begin{equation}
 \Pr \{ \text{a spike in $[t, t+dt )$} \} = f(V(t)) dt,
  \label{100520_19Aug16}
\end{equation}
where $f(\cdot)$ is a nonnegative intensity function.
%If $f(V) \propto \exp(-\beta (V-V_{th}))$ and $\beta \to \infty$, the soft-threshold model is equivalent to the hard-threshold model.

When $\sigma_{i}(t)=0$ in Eq.~\eqref{134035_28Oct15},
the solution is given by
\begin{equation}
 V_{i}(t) = V_{i}(0) e^{-g_{i} t}
  + \sum_{j} \int_{-\infty}^{t} \exp(-g_{i}(t-s)) I_{ij}(t) ds.
  \label{095415_19Aug16}
\end{equation}
If we assume that
$\kappa_{ij}(\cdot)$ is the dirac's delta function and
that $t$ is much greater than $1/g_{i}$.
Then, Eq.~\eqref{095415_19Aug16} can be approximated by
\begin{equation}
 V_{i}(t) =
  \sum_{j} \sum_{\{ f:t_{j}^{(f)}<t \}}\ \exp(-g_{i}(t-t_{j}^{(f)})).
  \label{100458_19Aug16}
\end{equation}
By choosing $f(V) \propto \exp(-\beta (V-V_{th}))$ and discretizing
the time so that $dt=1$,
the combination of Eqs.~\eqref{100458_19Aug16} and
\eqref{100520_19Aug16} reduces to the GLM \eqref{101719_19Aug16}
with $A_{i0} = - \beta V_{th}$, $A_{ij}(k) = \beta \exp(-g_{i} k)$,
and exponential inverse link function $\phi(\cdot)=\exp(\cdot)$.
For more detailed discussion, please refer to~\citep{gerstner2002spiking,Gerstner2014nd,Paninski2007493}.

\subsubsection{Network likelihood models}
\label{section:NLM}

Network likelihood models
(NLMs)~\citep{Okatan2005nc,Kim2011ploscb,Stevenson2009ieee} are often
adopted as a generative model for spike-train data in a continuous-time domain.
Let $N_{i}(t)$ be the total number of the $i$th neuron's spikes observed
before time $t$. In NLMs, it is assumed that $N_{i}(t)$ follows
an inhomogeneous Poisson process 
$Poisson(\cdot|\lambda_{i}(t))$
with the conditional intensity function
\begin{equation}
 \lambda_{i}(t) = \exp
  \left(
   A_{i0} + \sum_{j=1}^{P} \sum_{k=1}^{K} A_{ij}(k) I_{jk}(t)
  \right),
\end{equation}
where
\begin{equation}
 I_{jk}(t) = \int_{0}^{t} \xi_{k}(t-s) N_{j'}(s) ds
\end{equation}
is a convolution of a linear filter $\xi_{k}(\cdot)$ and a spiking
history of the $j^{th}$ neuron.
A typical example of the linear filter is a rectangular windows of
duration $W$ defined as
\begin{equation}
 \xi_{k}(u) =
  \begin{cases}
   1 & \text{(if $u \in [(k-1)W, kW)$)}
   \\
   0 & \text{(otherwise)}
  \end{cases}.
\end{equation}
NLMs can be converted into GLMs if the spiking history
$N_{i}(t)$ is discretized into bins with a fixed window $W$, and $W$ is
so small that any bin has at most a single spike.

\subsubsection{Calcium fluorescence model}
\label{section:Calcium}

When $x_{i}(t)$ is a continuous variable indicating the intensity of
calcium fluorescence, we have to consider the fact that the
fluorescence signal has a fast initial rise upon an action potential followed by a slow decay.
A standard model to reflect this feature is as
follows~\citep{Mishchenko2011aas}:
\begin{align}
 z_{i}(t+1) &= (1-\alpha_{i}) z_{i}(t) + s_{i}(t)
 \label{091002_1Oct15}
 \\
 x_{i}(t) &= a_{i} z_{i}(t) + b_{i} + \zeta_{i}(t),\quad
 \zeta_{i}(t) \sim \mathcal{N}(0,\tau),
 \label{091009_1Oct15}
\end{align}
where $z_{i}(t)$ denotes the intracellular calcium concentration
and $s_{i}(t)$ is a spike indicator variable such that $s_{i}(t) = 1$ if the
$i^{th}$ neuron fires at time $t$ and $s_{i}(t)=0$ otherwise.
$\theta = \left\{ \alpha_{i}, a_{i}, b_{i}, \tau_{i} | i=1,\cdots,
P \right\}$ comprises the parameters of the model.
Nonlinearities such as saturation can also be modeled as proposed
by~\cite{Vogelstein2009bpj}.
In general, we cannot directly observe the variable $s_{i}(t)$ in this
setting. To cope with this issue, Eqs.~\eqref{091002_1Oct15} and
\eqref{091009_1Oct15} are often combined with
Eqs.~\eqref{091137_1Oct15} and \eqref{091141_1Oct15} in which
$x_{i}(t)$ is replaced by $s_{i}(t)$.

\subsubsection{Hawkes process model}
\label{sec:hawkes}

\cite{linderman2014discovering,linderman2015scalable} propose a Hawkes multivariate process to model a network of purely excitatory neurons. 
The model is considered as a generalized linear model (section \ref{section:glm}) with Poisson observations and a linear link function. 
The network model is defines by a binary matrix that indicates the existence of a directed connection between two neurons, a second weight matrix to represent the strength of each connection, and a vector specifying the transmission delay distribution for each directed connection.

\cite{linderman2015scalable} enhanced the computational performance of their former approach \citep{linderman2014discovering} with a new discrete time model to assume that neurons do not interact faster than an arbitrary time unit and a variational approximation of the former Gibbs sampling solution in order to make the model fully conjugate.

\subsubsection{Dynamic Bayesian Network}
\label{section:dbn}

Dynamic Bayesian Networks (DBN) \citep{murphy2002dynamic} extend Bayesian Networks for time-series modeling. 
A DBN is usually defined as a directed acyclic graph where nodes represent random variables at particular times, and edges indicate a conditional probability dependence  $P(x_{i,t}|x_{j,t-k})$ where where $x_{i,t}$ denotes the $i^{th}$
neuron's activity at that time $t$. 

\cite{eldawlatly2010use} demonstrated the feasibility of using DBM to infer the effective connectivity between spiking cortical neurons simulated with a generalized linear model (see section \ref{section:glm}). They use a simulated annealing algorithm to search over the space of network structures and conditional probability parameters. \cite{patnaik2011discovering} present a significantly faster fitting algorithm that consists of identifying, for each node, parent-sets with mutual information higher than a user-defined threshold. An upper bound of the Kullback-Leibler between the inferred distribution and the true distribution is defined as a function of the user-defined threshold.  

\subsubsection{Maximum entropy model}
\label{section:maxentropy}

The maximum-entropy model \citep{yeh2010maximum,roudi2015multi} assumes that the network state probability distribution is given by an exponential function of the network energy $E[S_i]$ = $E[(x_1,...,x_n)]$ such that the entropy is maximized while satisfying any statistical constraints. 
When the first and second-order statistics are given, the state probability distribution is given by 
\begin{equation}
  P(x_1,...,x_n) = \frac{1}{Z} \exp(-E[(x_1,...,x_n)])
  = \frac{1}{Z} \exp( \sum_i b_i x_i + \sum_{i \neq j} J_{ij} x_i x_j).
\end{equation}
This is an extension of the Ising model \citep{McCoy:2010} with spatial  connections potentially occurring between any neurons as well as temporal correlations.

The main limitation of the maximum entropy model is its computational complexity. Recent studies demonstrated its applicability to a few tens of neurons using only first and second-order statistics without temporal correlations \citep{yeh2010maximum}.

\subsection{Estimation of Model Parameters}

\subsubsection{Maximum likelihood method}
\label{MLE}

The standard method to determine the model parameters is the maximum
likelihood (ML) method. The likelihood of a parameter vector $\theta$ given a data set $D$ is defined as $p(D|\theta)$, the probability of reproducing the data. 
Here we denote denotes the negative log-likelihood function as
$J(\theta|D) = -\log p(D|\theta)$.
In the ML method, the
parameters are chosened such that
\begin{align*}
 \theta^{\ast} &= \mathop{\text{argmin}}_{\theta} J(\theta|D).
\end{align*}

When the generative model uses a Gaussian distribution for observed data, the ML method reduces to {\itshape the least square method}.
For example, in the case of the AR model (\eqref{ARmodel}), maximization of the log likelihood
is equivalent to minimization of {\itshape the sum of squared residuals}
\begin{equation}
 J(\theta|D) = \sum_{i=1}^{P} \sum_{t=1}^{T} \epsilon_{i}^{2}(t)
  = \sum_{i=1}^{P} \sum_{t=K+1}^{T} 
  \left[
   x_{i}(t) - A_{i0} -
   \sum_{j=1}^{P} \sum_{k=1}^{K} A_{ij}(k) x_{j}(t-k)
  \right]^{2},
\end{equation}
where $D \equiv
\left\{ x_{i}(t) | i=1,\dots, P; t=1,\dots, T\right\}$ denotes the observed data set.

This optimization can be achieved analytically in the case of least square problem, or in general by iterative optimization algorithms
such as the gradient descent methods.
When the model is represented using a set of hidden variables,
a standard way is to use the expectation-maximization (EM) algorithms.

\subsubsection{Regularization and Bayesian inference}
\label{regularizer}

A disadvantage of the ML method is that it often suffers from overfitting when the number of parameters is large relative to
the amount of data. A common way to deal with this issue is to introduce regularization to the parameters. 
From Bayesian statistical viewpoint, it can be considered as assuming a prior distribution for the parameters. This makes explicit the assumption for connection inference, which is necessary due to the ill-posed nature of the inverse problem.

The objective function with a regularization term is given as
\begin{align*}
 \theta^{\ast} &= \mathop{\text{argmin}}_{\theta}
 \left\{
 J(\theta|D) + \lambda \mathcal{R}(\theta)
 \right\},
\end{align*}
where $\mathcal{R}(\cdot)$ is a non-negative function and $\lambda$ is
a constant that controls the strength of regularization.
The most common regularizer is $L2$-norm regularizer, or the Ridge regularizer,
\begin{align}
 \mathcal{R}(\theta) = \sum_{r} \theta_{r}^{2},
\end{align}
where $r$ indexes the element of the set of parameters.
In the Bayesian framework, this is equivalent to assuming a Gaussian prior distribution for the parameters.
Another common regularizer is the $L1$-norm regularizer, or least absolute shrinkage and selection operator (LASSO) regularizer,
\begin{align}
 \mathcal{R}(\theta) = \sum_{r} |\theta_{r}|,
\end{align}
which favors sparse solution with many parameters being zero.
From Bayesian viewpoint, this is equivalent to assuming a Laplace (exponential in absolute value) prior distribution.

%From the point of view of statistics, regularizations are closely
%related to the maximum a posteriori (MAP) estimation in a Bayesian
%sense. In fact, the optimal parameters with the Ridge
%regularizer and the LASSO regularizer are equivalent to the MAP
%estimator with Gaussian and Laplace prior distributions of $\theta$,
%respectively.

Beside introducing regularization, Bayesian inference has several advantages
over the ML method. 
For example, if the model has hidden
variables, Bayesian inference is often used for estimating them as well
as the model parameters~\citep{Mishchenko2011aas}.
Also, the reliability of each value in
the parameter space can be evaluated as the posterior distribution.
The high density region of the posterior distribution is called the Bayesian
credible region, which can be used as an alternate to the confidence
interval in the statistical test~\citep{Bowman2008}.
Another advantage is that the marginal
likelihood, defined by the Bayes' theorem, can be used for
a criterion to select the best of several possible
models~\citep{Friston2011} (Section \ref{validation}).

\subsubsection{Approximate Bayesian inference methods}

While Bayesian inference offers favorable features as above, 
exact computation for the posterior distribution and the
marginal likelihood is often analytically intractable.
Thus it is it in practice important to select an approximation method. 
In the following, we briefly review
approximation methods for the Bayesian inference.
%\footnote{%
Details of their derivations and algorithms is beyond the scope of this paper and can be found in other articles \citep{chen2013overview,Murphy2012ml}.
\begin{enumerate}
 \item {\itshape Monte Carlo Sampling} approximates the
       target distribution (or value) as an aggregation (or average) of
       the finite number of random samples.
       Gibbs sampling~\citep{Casella1992},
       Metropolis-Hastings~\citep{Hastings1970} and sequential
       importance re/sampling~\citep{Liu1998} algorithms are typical
       examples of this class.
 \item {\itshape Laplace Approximation}~\citep{Raftery1996} is a deterministic
       approximation applicable to cases in which the MAP estimator
       can be easily obtained. The central idea is to approximate the
       target distribution as a multivariate Gaussian distribution using
       the Hessian matrix (the second-order partial derivatives) of the
       logarithm of the posterior distribution; thus, the Laplace
       approximation is not suitable for cases in which the posterior
       distribution is multi-modal or asymmetric.
 \item {\itshape Variational Bayes}~\citep{Attias1999} is another
       deterministic approximation method derived from the variational
       mean-field theory in statistical mechanics.
       Let us consider a case in which
       we want to obtain the posterior distribution
       $p(Z|D)$, where $Z=(z_{1},\dots,z_{m})$ is a set
       of all unknown variables (i.e. hidden variables and model
       parameters) such that $z_{i}$ ($i=1,\dots,m$) is a partition of
       $Z$. Also, consider the probability distribution family of $Z$
       that can be factored as $q(Z) = \prod_{i=1}^{M} q(z_{i})$.
       In the variational Bayes method, we approximate $p(Z|D)$
       by optimizing $q(Z)$ to minimize the KL divergence between $q(Z)$ and
       $p(Z|D)$.
       Optimization is achieved by an algorithm similar to the expectation and maximization (EM) algorithm.
       Variational Bayes is suitable for cases in which the joint distribution $p(D,Z)$
       is an instance of the exponential families~\citep{Wainwright2007}.
\end{enumerate}

\section{Case Study}

In this section, we review several examples of connectivity inference study, from both model-free and model-based approaches. We focus on the challenges addressed by each of the studies at each stage of the data processing pipeline (Figure~\ref{fig:pipe}).

\subsection{Model-Free Approaches}
\label{modelfree_studies}

The First Neural Connectomics Challenge \citep{guyon2014design}  encouraged the development of very diverse solutions. Remarkably, the top three solutions were all model-free methods. The organizers provided simulated neural network datasets for training and testing, as well as key references about the problem and sample code to get started. Neural network datasets were simulated using leaky integrate and fire neurons with short term synaptic depression implemented in the NEST simulator and the spike trains were transformed to fluorescence time-series using a fluorescence response model of calcium markers inside neurons. Each dataset consisted of one-hour time-series of neural activities obtained from fluorescence signals sampled at $20ms$ intervals with values normalized in the interval $[0,1]$. Organizers also provided information about the position of each neuron in an area of $1 \, mm^{2}$ and inter-neuron connectivity labels (i.e. excitatory connection or no connection). A validation dataset was provided for performance comparison by the Area Under the ROC Curve (AUC; see section \ref{validation}). Finally, the score on a separated test dataset was used for the final ranking of teams.  

\begin{figure}[h!]
 \centering
  %\reflectbox{%
  \includegraphics[width=0.85\textwidth]{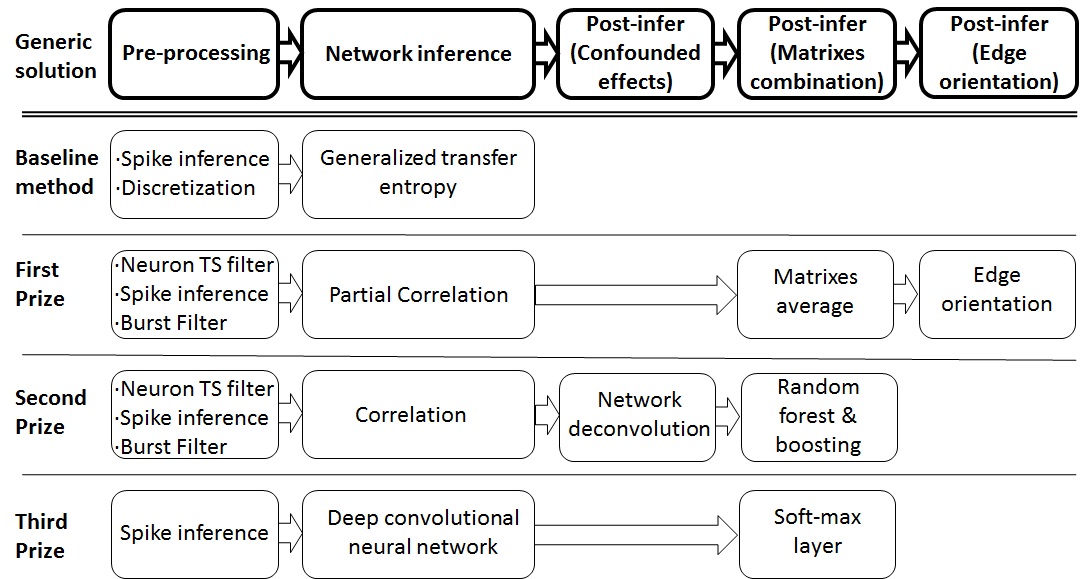}
  {\caption{Summary of top solutions of the First Neural Connectomics Challenge. The baseline method and the three winning methods are outlined according to their pre-processing, main connectivity inference, and post-processing steps. Note that not all methods implement all steps.}
  \label{fig:summarysolutions}}
\end{figure}

The three best solutions and the baseline method provided by the challenge organizers all consisted of several components: data pre-processing, main connectivity inference, and post-processing steps as discussed in section \ref{pipeline} (Figure~\ref{fig:summarysolutions} ). Below we review these top-ranked methods.

\subsubsection{Partial-correlation and averaging}
\label{modelfree_studies1}

The first prize solution \citep{sutera} achieved a performance of 0.94161 (AUC score). 
Its pre-processing consisted of four steps: i) low-pass filtering to remove noise, excessive fluctuations, and light scattering artifacts, ii) backward differentiation for spike detection, iii) hard thresholding to eliminate small values below a positive parameter $\tau$, and iv) weighting the importance of spikes as a function of the inverse of network overall activity. The motivation for this last step was to eliminate the effects of synchronized bursting periods,  much as generalized transfer entropy does (see section \ref{informationtheoreticmethods}).

The main connectivity inference employed the Partial-Correlation (PC) matrix (Section \ref{statisticalmethods}), using the real valued signal computed in the previous pre-processing steps. The post-processing consisted of two steps: i) Several PC matrices were averaged according to different $\tau\,'s$ and different low pass filters to increase the robustness of the solution. ii) The symmetric average PC matrix was transformed into a directional indicator by multiplying each edge by an orientation weight computed according to the average activation delay between each neuron pair.

The main difficulty in scaling this solution to larger neural networks is the computation burden for calculating the inverse matrices. As an example, the computation of the challenge solution connection matrix with 1000 neurons took 30 hours on a 3 $GHz$ i7 desktop PC with 7 $GB$ of RAM.

\subsubsection{Matrix deconvolution and adaptive averaging}
\label{modelfree_studies2}

The major features of the second prize solution \citep{magrans} are that it used matrix deconvolution for eliminating indirect connections and introduced learning to optimally combine several connectivity matrices rather than just using simple averaging.
The pre-processing pipeline consisted of three steps: i) spike train inference based on the OOPSI algorithm \citep{vogelstein2009oopsi}, ii) hard thresholding according to a parameter value $\tau$, as in the first prize solution, and iii) removal of time segments with overall network activity above a given threshold $\theta$.

Connectivity inference employed Pearson correlation between the real valued signal computed from the pre-processing. 
A network deconvolution algorithm \citep{feizi2013network} was then used to eliminate the combined effect of indirect paths.
If the effect of direct connection is given by matrix $W_d$, under the assumption of linearity, the combined effects of all the direct and indirect connections follows
\begin{equation}
  W_o = W_d + W_d^2 + W_d^3 + ... = W_d (I - W_d)^{-1}
\label{eq:convolution}
\end{equation}
From this relationship, the matrix deconvolution method estimates the direct connection matrix $W_d$ from the observed connections matrix $W_o$ by
\begin{equation}
  W_d = W_o (I + W_o)^{-1}
\label{eq:deconvolution}
\end{equation}
Finally, the connectivity matrices computed according to different values of $\tau$ and $\theta$ are combined, but unlike the first prize solution, using a function learned with a non-linear regression method for optimal performance.

Computation of the challenge solution took nearly 48 hours on a 3 $GHz$ i7 desktop PC with 32 GB of RAM. The most significant limitation of this solutions is its high computational cost. It does not try to infer self-connections. It does not try to identify the causal direction. It assumes that both training and testing networks have similar statistical properties.
  
\subsubsection{Convolutional neural network approach}
\label{modelfree_studies3}

The third prize solution \citep{lukasz} proposed to automatically extract features from binary spike train pairs using a Convolutional Neural Network (CNN). Pre-processing consisted of three steps: i) spike train inference using discrete differentiation, ii) hard thresholding as in the first prize solution, and iii) spike train normalization in [0,1]. 
Feature extraction was done with a CNN based on Lenet5  \citep{lecun1998gradient} followed by a softmax layer that produced binary output in the event of a connection. This method was not designed to evaluate directions of connections. 

Input data consisted of binary matrices of 3 by 330 where rows 1 and 2 corresponded to segments of the spike trains of two neurons. Row 3 contained information about the overall network activity, making it possible to identify synchronous bursting episodes. Network training was done with gradient descent. 

The computational cost for this solution is very high. For instance, the challenge solution took more than 3 days on 8 server machines working in parallel, where each machine was equipped with 32 GB of RAM and GPU unit with 2496 cores. In addition to the extremely high computational cost, this solution has the same limitations than the second prize solution. 

After the Connectomics Challenge, \cite{veeriah2015deep} further advanced the CNN approach by integrating both the pre-processing step of spike train inference and connectivity inference into a single neural network architecture to surpass the first prize performance using the same dataset.
Its architecture consisted of two sub-networks and a final classification layer: i) a convolutional neural network with max-pooling layers responsible for identifying relevant shapes in the fluorescence time-series with tolerance for minor time translations, ii) a recurrent neural network (RNN) to model temporal sequences of relevant events. These are duplicated to capture the features of each neuron pair. Finally, a dynamically programmed layer aligned the RNN outputs and computed a connection probability for each neuron pair. Remarkably, this method does not require a separate pre-processing step to handle the synchronized bursting phenomenon.

Although the concept of an end-to-end artificial neural network is appealing, the performance comparison in \cite{veeriah2015deep} deserves further consideration.
Their reconstructed performances for generalized transfer entropy and partial correlation methods are lower than the performance documented during the First Neural Connectomics Challenge using the same data sets \citep{guyon2014design, orlandifirst, sutera}. 

While these CNN-based approaches presented good performance with simulated data, it remains to see if such classifiers generalize well to real data, for which ground truth training data are rarely available.
%Both the artificial neural network based solutions discussed in this section and the second prize solution discussed in section \ref{modelfree_studies2} require training data consisting of the ground truth connectivity matrix and the activity recordings from each neuron. This training data is necessary to fit a function able to predict the probability of two neurons being connected given their activity recordings or derived features vectors. Although these approaches are very successful with simulated data, it is very hard to apply them to real data because they have the fundamental limitation to assume that we have access to the ground truth connectivity structure of a training data set that has similar properties than the network that we want to reconstruct. 

\subsubsection{Other model-free approaches}
\label{other_modelfree_studies} 

Besides the main results of the First Neural Connectomics Challenge, several \textit{in vivo} and \textit{in vitro} studies rely on model-free methods.  
\cite{fujisawa2008behavior} use silicon microelectrode arrays for \textit{in vivo} recording from layers 2,3 and 5 of the medial prefrontal cortex in rats during a working memory task. Connectivity inference is performed by identifying sharp spikes or troughs in the cross-correlograms. The connectivity significance is assessed by first creating several sets of slightly perturbed spike trains and then computing the statistical significance of the original cross-correlograms with respect to those constructed with the perturbed data.
   
\citep{cohen2011measuring} discuss recent surge in cortical processing studies enabled by recording technologies like microelectrode arrrays and calcium imaging (see section \ref{realdatasource}). They summarize possible causes of discrepant findings in several correlation based studies due to technical reasons like poor spike sorting; experimental factors like different external stimulus, different time bins and/or spike train durations or confounded effects due to non-stationary internal states. They suggest that multivariate point process models provide a more complete and statistically principled approach to cope with the challenges of inferring the connectivity structure.   

\subsection{Model-Based Approaches}

\subsubsection{Generalized linear models}

Pillow and colleagues applied GLM frameworks for a stochastic integrate-and-fire neuron model \citep{Paninski2004nc} to the data of a complete neural recording from ganglion cells in a retina slice \citep{Pillow2008n}. They could construct a model including both the stimulus response and cross-connections of all neurons and showed through model-based analysis that correlated firing can provide additional sensory information.

Stevenson and colleagues derived a Bayesian inference framework for a spiking neuron model with arbitrary synaptic response kernels \citep{Stevenson2009ieee}. They applied the method to neural recording data from monkey motor cortex and used the log likelihood and correlation coefficient criteria for cross validation. They further used infinite relational model clustering \citep{kemp2006learning} to detect cluster structures among recorded neurons.

While \cite{Stevenson2009ieee} used arbitrary synaptic kernel functions with discrete time bins, Song and colleagues \citep{Song2013jcn} proposed the use of a set of kernel functions, such as B-splines, with which smoothness of kernel functions are built-in with a relatively small number of parameters.

\cite{oba2016empirical} propose a GLM framework combined with empirical Bayesian testing. Null samples are created by time-shifting real spike trains for a sufficiently large time lag. They show an improved computational performance without decreasing inference performance on a simulated neural network. They also apply this method to real calcium imaging data from a cultured slice of the CA3 region of a rat hippocampus.

\subsection{Calcium fluorescence model}

In applying a connection estimation method to calcium fluorescent imaging data, a basic way is to estimate spike trains from fluorescence waveforms by spike deconvolution and then to apply a spike-based connectivity inference algorithm. \cite{Mishchenko2011aas} proposed a framework combining both steps of spike estimation and connection estimation into a unified stochastic generative model. 
\cite{fletcher2014scalable} proposed a computationally less expensive variant of this approach.

\section{Discussion}
\label{discussion}

\subsection{Challenges and Solutions}
\label{solutions}

While model-free methods usually address pre- and post-processing separately from connectivity inference, model-based methods often address issues such as noise and connection directionality in the main inference process by incorporating those factors into generative models.

The distinction between apparent and real connections, as well as their directionality, is incorporated into neural dynamics equations in model-based methods. Weights most consistent with the data are selected by the maximum likelihood or maximum posterior probability criterion.
Among model-free methods, transfer entropy methods address connection directionality directly during the inference, while removal of apparent connections can be addressed by matrix deconvolution in post-processing \citep{magrans}. Partial-Correlation can also remove aparent connections but is not able to identify the causal direction. \cite{sutera} proposes a post-processing step to identify the causal direction. The limited temporal resolution of calcium imaging impose severe performance limitations to these methods. Better solutions to the problem of identifying the direction and inferring apparent connections should involve improving both algorithms and instrumentation. Table \ref{tab:solution} summarizes the solutions that were devised to overcome challenges along the data processing pipeline.

\renewcommand{\chprefix}{3}
\renewcommand\thetable{\chprefix}
\begin{table}[h]
  \caption{Summary matrix of the solutions that were devised to overcome each of the challenges. We can observe that each solution was implemented either as a pre/pro-processing step or within the main inference method. }
  \label{tab:solution}
  %\footnotesize
  \scriptsize
  %\tiny
  \centering
  \hspace*{-1cm}
  \begin{tabular}{|K{2cm}|K{3cm}|K{3cm}|K{3cm}|K{3cm}|}
	\hline
	Challenges & Pre-processing & \multicolumn{2}{c|}{Connection inference} & Post-processing \\
	\cline{3-4}
	& & Model-free & Model-based & \\
	\hline
	
	Apparent connection &
	&
	\cite{sutera,veeriah2015deep} &
	\cite{Pillow2008n,Stevenson2009ieee,
	Mishchenko2011aas,fletcher2014scalable} &
	\cite{magrans} \\
	\hline
	
	Directionality &
	&
	\cite{stetter2012model,garofalo2009evaluation,
	ito2011extending,veeriah2015deep} &
	\cite{Pillow2008n,Stevenson2009ieee,
	Mishchenko2011aas,fletcher2014scalable} &
	\cite{sutera} \\
	\hline
	
	Cellular diversity &
	&
	&
	&
	\\
	\hline
	
	Synaptic diversity &
	&
	&
	\cite{Pillow2008n,Stevenson2009ieee,
	Mishchenko2011aas,fletcher2014scalable} &
	\\
	\hline
	
	Non-stationarity &
	\cite{cohen2011measuring,sutera,magrans,garofalo2009evaluation}&
	\cite{wollstadt2014efficient,stetter2012model,veeriah2015deep} &
	\cite{linderman2014framework} &
	\\
	\hline
	
	Noise &
	\cite{cohen2011measuring,sutera,magrans,wollstadt2014efficient,
	stetter2012model,garofalo2009evaluation} &
	\cite{veeriah2015deep} &
	\cite{Pillow2008n,Stevenson2009ieee,
	Mishchenko2011aas,fletcher2014scalable} &
	\\
	\hline
	
	Time/Space resolution &
	&
	\cite{veeriah2015deep} &
	\cite{Mishchenko2011aas,fletcher2014scalable} &
	\\
	\hline
	
	Hidden neuron/ External input &
	&
	&
	\cite{kulkarni2007common,vidne2012modeling,rezende2011variational} &
	\\
	\hline
	
	Prior knowledge &
	&
	&
	\cite{Pillow2008n,Stevenson2009ieee,
	Mishchenko2011aas,fletcher2014scalable} &
	\\
	\hline
	
	Scalability &
	&
	\cite{sutera,magrans} &
	\cite{fletcher2014scalable} &
	\\
	\hline
	
  \end{tabular}
\end{table}

%\begin{table}[h!]
  %\centering
  %\includegraphics[width=\textwidth]{challenge_matrix.pdf}
  %{\caption{Summary matrix of the solutions that were devised to overcome each of the challenges. We can observe that each solution was implemented either as a pre/pro-processing step or within the main inference method. }
  %\label{fig:matrix}}
%\end{table}

%FloatBarrier

\subsubsection{Non-stationarity}

Modeling synaptic plasticity is a key to understand how memory and learning mechanisms are realized in neural circuits. Recent solutions to this challenge incorporate synaptic dynamics into a GLM framework and improved scalability within Bayesian \citep{linderman2014framework} and convex optimization settings \citep{stevenson2011inferring}. Using a model-free approach to understand the evolution of synaptic weights, \cite{wollstadt2014efficient} proposed a transfer entropy estimator that requires an ensemble of spike trains from equivalent and independent experiments.

Synaptic plasticity is not the sole source of non-stationarity. Time periods when a large percentage of neurons fire at a high rate is an additional non-stationary phenomenon and a source of confusion for most network inference methods. 
Generalized transfer entropy (Section \ref{section:gte}), 
screens out the use of non-informative time periods. 
A similar selection could be developed for other model-free methods. 

In section \ref{modelfree_studies} we also discussed how winning solutions of the First Neural Connectomics Challenge chose a modular approach with a dedicated pre-processing step to remove non-informative time periods before applying different connectivity indicators. We further discussed a deep neural network solution by \citep{veeriah2015deep} that surprisingly, despite the lack of any pre-processing step or any specific design feature to remove high-rate time periods, is able to achieve improved performance with respect to the winning solution. A generalization of the above bi-modal switching behavior can also be accommodated with a model-based approach using a hidden Markov model \citep{lindermandiscovering}.

\subsubsection{Architecture}

Pre/post-processing methods have the potential to improve the network reconstruction performance across different methods. For instance, modular architectures like the top solutions of the connectomics challenge (Section \ref{modelfree_studies}) implement a number of pre-processing steps to reduce noise and to infer spike trains.  However, in applying such modular approaches, care has to be taken not to lose valuable information about the underlying direct connectivity and the directionality.
% due to required trade-offs at each step.
On the other side of the spectrum, the approach using a deep neural network \citep{veeriah2015deep} is unique in providing solutions for multiple pre-processing and post-processing steps within a coherent model-free architecture. 

In the model-based approach, \cite{Mishchenko2011aas} and later \cite{fletcher2014scalable} employed time-series of neural activities obtained from fluorescence signals as inputs, while the spike times at an arbitrary sampling rate were additional latent variables. Table \ref{tab:solution} summarizes the solutions that were devised to overcome each of the challenges. We can observe that each solution was implemented either as a pre/pro-processing step or within the main inference method.   
%Identification and estimation of hidden neurons and/or unobservable external inputs remain as one of the most significant open issues; we address those in section \ref{hidden_neurons}.

\subsubsection{Scalability}

A major drawback to model-based methods is their scalability. The more sophisticated the model, the more parameters need to be estimated, which requires more data sampled from a stationary distribution. Prior knowledge, such as sparseness in connections, can be addressed in model-based methods by assuming a prior distribution in the model parameters and by applying Bayesian inference.
Introduction of a sparseness prior, or an equivalent regularization term, can make inference from smaller samples more reliable, but this often increases the computational burden. %%Model-free methods based on correlation are generally not computationally intensive, so they can easily be scaled to a network of thousands of neurons. 

\subsection{Hidden Neurons}
\label{hidden_neurons}

While neural recording methods are progressing rapidly allowing whole-brain imaging with cellular resolution in smalls animal like \textit{Caenorhabditis elegans} \citep{nguyen2016whole} and zebra fish \citep{ahrens2013whole}, monitoring all neurons in the brain is still not possible. In typical calcium fluorescent imaging, neurons in only one section of a cortical column are recorded. Neglecting hidden neurons within the network, or unobservable external inputs, can infer erroneous connections. Methods for addressing this hidden input problem are still in early stages of development \citep{roudi2015multi}.

The issue of hidden nodes, however, is not the only problem in neural connectivity inference. Similar problems have been addressed in gene regulation and signaling cascade networks, for examples, and methods that have been developed in other fields of computational biology and network science may provide helpful intuition and guidance \citep{Su2014sr}.

The challenge of hidden nodes groups different application scenarios under the same name, but the common denominator here is a generative model that consists of observed and hidden components. The GLM (Section \ref{section:glm}) and maximum entropy settings (Section \ref{section:maxentropy}) are the most common approaches to describe the observed neurons. However, the key differential contribution in all cases is the hidden neurons model. A single, latent process can model the hidden inputs as random effects, enhancing network inference among observed neurons by avoiding false connections due to common hidden inputs \citep{kulkarni2007common, vidne2012modeling}. A more sophisticated group of contributions proposes multiple latent processes, for instance to study the feasibility of biophysically plausible hypotheses about how multi-level neural circuits are able to learn and express complex sequences \citep{rezende2011variational}. Switching behavior, described in the previous section, can also be seen as a special case of the hidden nodes challenge.

\subsection{Incorporating Prior Knowledge}
\label{prior}

Detailed anatomical synaptic maps \citep{yook2013mapping}, and connectivity maps between cell types across cortical layers \citep{potjans2014cell} are  valuable sources of prior information that we should exploit to improve both inference and computational performance. A straightforward  model-based approach would be to exploit anatomical prior information using more sophisticated regularizers like the adaptive elastic net \citep{zou2009adaptive}, by embedding anatomical information  in the adaptive weights instead of computing them using ordinary least squares \citep{wu2016regularized}. Prior information could also be incorporated as a post-processing method using graph sparsification algorithms to preserve certain graph theoretical properties \citep{ebbes2008sampling,lindner2015structure}.

A possible way to deal with hidden neurons when analyzing cortical microcircuits is to take advantage of the highly replicated structure across layers and cortical columns. A possible framework to implement this idea, proposed by \citep{kim2011network}, infers the parameters of a Kronecker model. In other words, a low-dimensional connectivity graph that approximately self-replicates across observed and the hidden parts. Within this application, the exploitation of prior knowledge could enhance both computational and inferential performance.

The network inference problem could also be considered a process of selecting for each neuron a sub-set of pre-synaptic input neurons. Therefore, it would be reasonable to explore the possibility of applying feature selection algorithms \citep{liu2012feature} to the problem at hand. From this point of view, multi-task feature selection algorithms \citep{obozinski2006multi,wang2016multi,zhou2010exclusive} may be another interesting way to take advantage of the highly replicated structure across layers and cortical columns. These algorithms propose clever ways to jointly learn features across several tasks, maximizing information sharing while minimizing negative transfer. Therefore, application of these methods has the potential to deliver superior inference and computational performance with respect to single-task learning approaches while using the same amount of data.  

A common assumption of many inference methods is a simplistic model structure that does not represent the true synaptic and cellular diversity of local neural microcircuits (see section \ref{biophysical_challenges}). Future research methods should aim for approaches able to fit more biophysically plausible models.
Recent wide-field calcium imaging of thousands of neurons over millimeters of brain tissue \citep{mohammed2016integrative}, the large number of latent parameters to model hidden neurons and non-stationarity, are all sources of increasing computational complexity and a strong impetus to continue improving the computational efficiency of network inference methods \citep{lee2016hawkes}.

\section{Conclusion}
\label{conclusion}

In this paper, we reviewed methods for inference of neural connectivity based on activity recordings from a large number of neurons. We first identified biophysical and technical challenges along the data processing pipeline and then formulated model-free and model-based approaches for the core process of connectivity inference. We further investigated in previous works how those challenges were addressed using what methods. As a result, we identified favorable methods issues that deserve further technical developments, most notably coping with hidden neurons.

Connectivity inference itself is an interesting and deep mathematical problem, but the goal of connectivity inference is not only to precisely estimate the connection weight matrix, but also to illustrate how neural circuits realize specific functions, such as sensory inference, motor control, and decision making, and to understand the base-line brain dynamics upon which those function would be based. If we can perfectly estimate network connections from anatomical and activity data, then computer simulation of the network model should be able to reproduce the function of the network as well as the resting-state dynamics. But given inevitable uncertainties in connectivity inference, reconstruction of functions in a purely data-driven way might be difficult. How to extract or infer a functional or computational network from a data-driven network, or even to combine known functional constraints as a prior for connectivity inference, is a possible direction of future research.

%Connectivity inference itself is an interesting and deep mathematical problem, but the goal of connectivity inference is not only to precisely estimate the connection weight matrix, but also to illustrate how neural circuits realize specific functions, such as sensory inference, motor control, and decision making. If we can perfectly estimate network connections from anatomical and activity data, then computer simulation of the network model should be able to reproduce the function of the network. But given inevitable uncertainties in connectivity inference, reconstruction of function in a purely data-driven way might be difficult. How to extract or infer a functional or computational network from a data-driven network, or even to combine known functional constraints as a prior for connectivity inference, is a possible direction of future research.

\section*{Acknowledgment}

This work was supported by the Program for Brain Mapping by Integrated Neurotechnologies for Disease Studies (Brain/MINDS) from Japan Agency for Medical Research and Development (AMED), and internal funding from the Okinawa Institute of Science and Technology Graduate University. We thank Steven Aird for editing the manuscript.

%\newcommand{\sectionbreak}{\clearpage}
%\appendix
%\section{Summary of model-free and model-based methods}

%\input{model_free_table}
%\input{model_base_table}

\section*{References}

\bibliographystyle{elsarticle-harv} 
\bibliography{ConnectionInference}

\end{document}